\documentclass[12pt]{article}
\usepackage{graphicx}

\newcommand{\mtwonui}{\mbox{$m^2(\nu_{\mathrm i})$}}

\newcommand{\mnui}{\mbox{$m(\nu_{\mathrm i})$}}

\newcommand{\mnue}{\mbox{$m(\nu_{\mathrm e} )$}}

\newcommand{\mtwonue}{\mbox{$m^2(\nu_{\mathrm e} )$}}

\newcommand{\mee}{\mbox{$m_\mathrm{ee}$}}

\newcommand{\mnu}{\mbox{$m_\nu$ }}
\newcommand{\ttwo}{\mbox{$\rm T_2$}}

\newcommand{\kr}{\mbox{$\rm ^{83m}Kr$}}
\newcommand{\rhenium}{\mbox{$\rm ^{187}Re$}}
\newcommand{\etot}{\mbox{$E_{\rm tot}$}}
\newcommand{\detot}{\mbox{$dE_{\rm tot}$}}

\newcommand{\gfermi}{\mbox{$G_{\rm F}$}}
\newcommand{\mlep}{\mbox{$M_{\rm lep}$}}
\newcommand{\mhad}{\mbox{$M_{\rm nucl}$}}
\newcommand{\ev}{\mbox{$\rm eV/c^2$}}
\newcommand{\evtwo}{\mbox{$\rm eV^2/c^4$}}
\newcommand{\bdec}{\mbox{$\beta$~decay}}
\newcommand{\bspec}{\mbox{$\beta$~spectrum}}
\newcommand{\belec}{\mbox{$\beta$~electron}}

\newcommand{\ezero}{\mbox{$E_0$}}

\newcommand{\mtwolep}{\mbox{$|M^2_{\rm lep}|$}}
\newcommand{\mtwohad}{\mbox{$|M^2_{\rm nucl}|$}}

\newcommand{\vj}{\mbox{$V_{\rm j}$}}
\newcommand{\erec}{\mbox{$E_{\rm rec}$}}
\newcommand{\be}{\begin{equation}}
\newcommand{\ee}{\end{equation}}
\newcommand{\bea}{\begin{eqnarray}}
\newcommand{\eea}{\end{eqnarray}}

\newcommand{\etal}{\mbox{\it et al.}}

\title{Direct determination of Neutrino Mass from $^3H$ $\beta$-spectrum}

\author{C. Weinheimer\\
\it \footnotesize
Institut f\"ur Kernphysik, Westf\"alische 
Wilhelms-Universit\"at M\"unster, 48149 M\"unster, Germany,\\ 
\it \footnotesize Email: weinheimer@uni-muenster.de
}

\begin{document}

\maketitle

\begin{abstract}
The investigation of the endpoint region of the tritium \bdec\ spectrum is still
the most sensitive direct method to determine the neutrino mass scale. 
In the nineties and the beginning of this century the tritium \bdec\ experiments at Mainz and Troitsk have reached a sensitivity
on the neutrino mass of 2~\ev . They were using a new type of high-resolution spectrometer with large sensitivity, the MAC-E-Filter, 
and were studying the systematics in detail. Currently, the KATRIN experiment is being
set up at Forschungszentrum Karlsruhe, Germany. KATRIN will improve the neutrino mass sensitivity by one order of magnitude down to 0.2~\ev , sufficient to cover the
degenerate neutrino mass scenarios and the cosmologically relevant neutrino mass range. 
\end{abstract}

\section{INTRODUCTION}
\begin{figure}[b!]
\includegraphics[width=\textwidth]{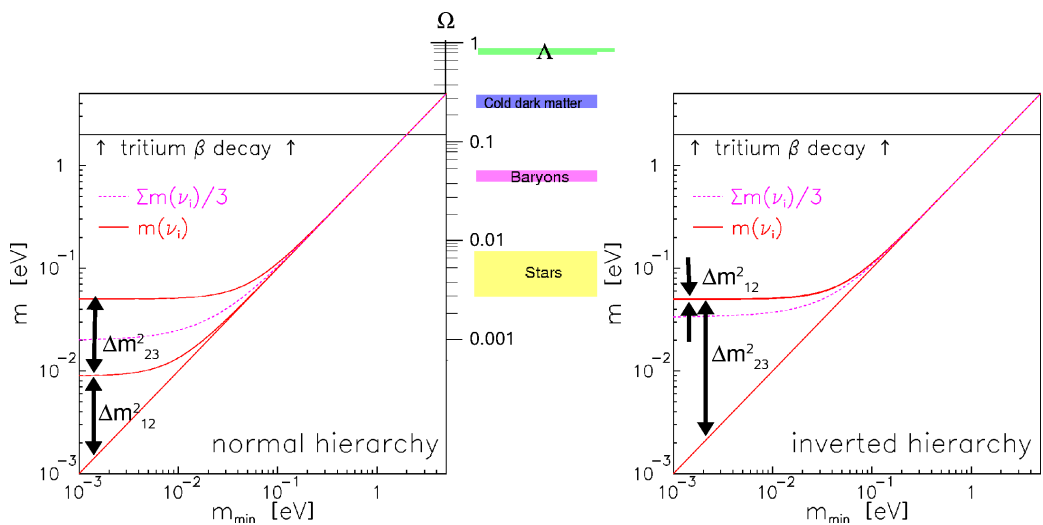}    
\caption{Neutrino mass eigenvalues
 \mnui\ (solid lines) and one third of the cosmologically relevant sum of the three neutrino mass eigenvalues $\sum m(\nu_i) /3$ (dashed line) 
as a function of the smallest neutrino mass eigenvalue $m_{\rm min}$ for normal hierarchy $m(\nu_3) > m(\nu_2) > m(\nu_1)$ 
(left) and inverted hierarchy  $m(\nu_2) > m(\nu_1) > m(\nu_3)$ (right).
The upper limit from the tritium \bdec\ experiments at Mainz and Troitsk 
on \mnue\  (solid line), which holds in the degenerate neutrino mass region for each 
\mnui , and for $\sum \mnui /3$ (dashed line) is also marked. The hot dark matter contribution $\Omega_\nu$ of the universe relating 
to the average neutrino mass $\sum \mnui /3$ is indicated by the right scale in the normal hierarchy plot and compared to all other known 
matter/energy contributions in the universe (middle). 
With the relic neutrino density of 336/cm$^3$ the laboratory neutrino mass limit from tritium \bdec\ $\mnue < 2 \ev$ 
corresponds to a maximum allowed neutrino matter contribution in the universe of $\Omega_\nu < 0.12$. 
\label{fig:nu_scenarios}} 
\end{figure}

We know from neutrino oscillation experiments that the
different neutrino flavors mix and can oscillate during flight from one flavor state into another. The
analysis of all neutrino oscillation experiments yields the mixing angles and the differences of squared neutrino mass eigenstates  
\cite{nu_osc}.
Clearly these findings prove that neutrinos have non-zero masses, but such ``interference experiments''
cannot probe the absolute mass scale. We have to parametrize our ignorance by a free parameter $m_{\rm min}$, the mass of the smallest
neutrino mass eigenstate (see fig. \ref{fig:nu_scenarios}).

\clearpage

The huge abundance of neutrinos left over in the universe from the big bang (336/cm$^3$) and their contribution to 
structure formation \cite{pastor_varenna}
as well as the key role of neutrino masses in finding the new Standard Model of particle physics \cite{nu_mass_theo_varenna1,nu_mass_theo_varenna2} 
make the absolute value of the neutrino mass one of the most urgent questions of astroparticle physics
and cosmology as well as of nuclear and particle physics.

\begin{figure}[tb]
\includegraphics[width=\textwidth]{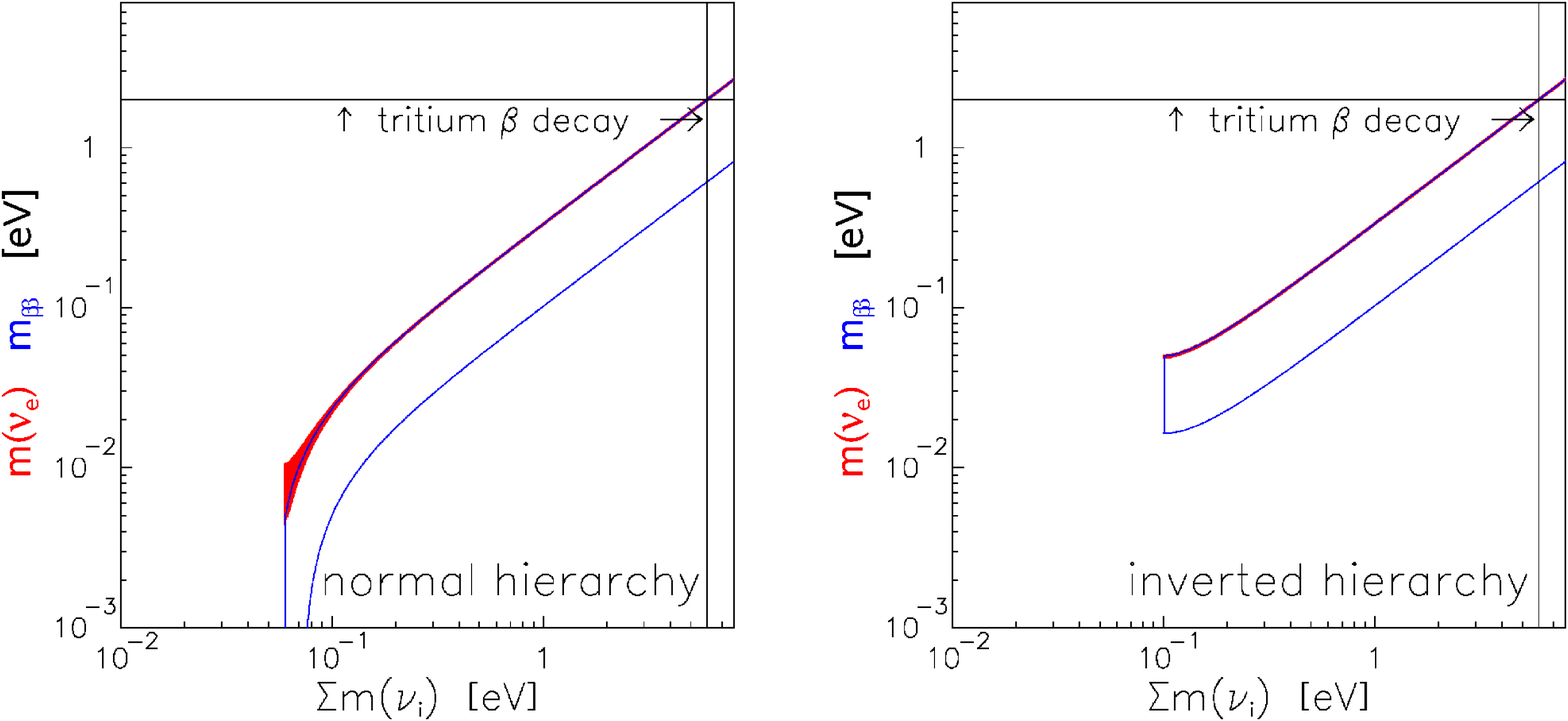}    
\caption{Observables of neutrinoless double \bdec\ \mee\ (open blue band) and of direct neutrino mass determination by single \bdec\ \mnue\ 
(red) versus the cosmologically relevant sum of neutrino mass eigenvalues 
$\sum \mnui$ for the case of normal hierarchy (left) and of inverted hierarchy (right). 
The width of the bands/areas is caused by the experimental uncertainties of the neutrino mixing angles \cite{pdg08}
and in the case of \mee\ also by the completely unknown Majorana- and CP-phases. 
Uncertainties of the nuclear matrix elements, which enter \mee , are not considered.
\label{fig:comparison_methods}} 
\end{figure}

There exist 3 different approaches to the absolute neutrino mass scale: 
\begin{itemize}
\item Cosmology\\
  Essentially the size of fluctuations is observed at different scales by using cosmic microwave background and
  large scale structure data. Since the light neutrinos would have smeared out fluctuations at small scales the power spectrum
  at small scales is sensitive to the neutrino mass. Up to now, only limits on the sum of the 3 neutrino masses have been obtained around
  $\sum \mnui < 0.61~\ev$ \cite{wmap2008}, which are to some extent 
model- and analysis dependent \cite{pastor_varenna,hannestad_bao97}.
\item Neutrinoless double \bdec ($0\nu\beta\beta$)\\
  A neutrinoless double {\bdec} (two \bdec s in the same nucleus at the same time with emission of two \belec s (positrons) while the (anti)neutrino
  emitted at one vertex is absorbed at the other vertex as a neutrino (antineutrino)) 
  is forbidden in the Standard Model of particle physics. It could exist, if the neutrino is its own antiparticle (``Majorana-neutrino'' in
  contrast to ``Dirac-neutrino''). Furthermore, a finite neutrino mass is 
  required in order to produce in the chirality-selective interaction a neutrino with a small component of opposite handedness on 
  which this neutrino exchange subsists. The decay rate would scale with the absolute square of the so called effective neutrino mass,
  which takes into account the neutrino mixing matrix $U$:
  \be\
    \Gamma_{0\nu\beta\beta} \propto \left| \sum U^2_{\rm ei} \mnui \right|^2 := \mee^2
  \ee
Here \mee\ represents the coherent sum of the \mnui-components of the $0\nu\beta\beta$-decay amplitudes and hence carries their relative phases
  (the usual CP-violating phase of an unitary $3 \times 3$ mixing matrix plus two so-called Majorana-phases). A significant additional uncertainty which
  enters the relation of \mee\ and the decay rate is the nuclear matrix element of the neutrinoless double \bdec . 
  There is one claim for evidence at $\mee \approx 0.4 ~\ev$ by part of the Heidelberg-Moscow collaboration \cite{klapdor04} and limits
  from different experiments in the 1~\ev\ range \cite{varenna_0nbb}.
\item Direct neutrino mass determination\\
  The direct neutrino mass determination is based purely on relativistic kinematics without further assumptions. 
  Therefore it is sensitive to the neutrino mass squared $m^2(\nu)$.
  In principle there are two methods: time-of-flight measurements and precision investigations of weak decays.
  The former requires very long baselines and therefore
  very strong sources, which only cataclysmic cosmological events like a core-collapse supernova could
  provide.  The non-observation of a dependence of the arrival time on energy of supernova neutrinos from SN1987a gave 
  an upper limit on the neutrino mass of 5.7~\ev\ \cite{pdg08}. 
  Unfortunately nearby supernova explosions are too rare
  and too little understood to allow an improvement into the sub-eV range.

  Therefore, aiming for this sensitivity, the investigation of the kinematics of weak decays and more explicitly the
  investigation of the endpoint region of a \bdec\ spectrum is still the most sensitive model-independent and direct method 
  to determine the neutrino mass. 
  Here the neutrino is not observed but the charged decay products are precisely measured. Using energy and momentum conservation 
  the neutrino mass can be obtained. In the case of the investigation of a \bspec\ usually the ``average electron neutrino mass''
  \mnue\ is determined (see sect. 2 below):
  \be\ \label{eq_define_mnue}
    \mnue^2 := \sum |U^2_{\rm ei}| \mnui^2
  \ee\ 
   This incoherent sum is not sensitive to phases of the neutrino mixing matrix.
\end{itemize}  
Figure \ref{fig:comparison_methods} demonstrates that the different methods are complementary to each other and 
compares them.

These lecture notes are structured as follows:
In section 2 the neutrino mass determination from the kinematics of tritium \bdec\ is described. 
Section 3 presents the previous tritium \bdec\ experiments, especially the experiments at Mainz and Troitsk. 
In section 4 an overview of the present KATRIN experiment
is given. 
This paper closes with a conclusion in section 5.
For a more detailed and complete overview on this subject we would like to refer to the reviews \cite{robertson,holzschuh,wilkerson,winter_book},
with \cite{otten08} being the most recent one.

\section{Neutrino mass from the tritium \bdec\ spectrum}

According Fermi's Golden Rule the decay rate for a \bdec\ is given by the square of the transition matrix element $M$ 
summed and integrated over
all possible discrete and continuous final states $f$  (from here on we use the convention $\hbar = 1 = c$ for simplicity):
\begin{equation}
\label{eq_fermisgoldenrule}
  \Gamma = 2\pi \sum \int | M^2 | df
\end{equation} 
Let us first calculate the density of the final states.
The number of different final states $dn$ of outgoing particles inside a normalization volume $V$
into the solid angle $d\Omega$ with momenta between $p$ and $p+dp$, 
or, respectively, with energies in the corresponding  
interval around the total energy \etot, is:
\begin{equation}
\label{eq_nstates}
  dn = \frac{V \cdot p^2 \cdot dp \cdot d\Omega}{h^3}
     = \frac{V \cdot p^2 \cdot dp \cdot d\Omega}{(2 \pi)^3}
     = \frac{V \cdot p \cdot \etot \cdot \detot \cdot d\Omega}{(2 \pi)^3} \quad .
\end{equation}
This gives a state density per energy interval and solid angle of
\begin{equation}
\label{eq_ndensity}
  \frac{dn}{\detot d\Omega} = \frac{V \cdot p \cdot \etot}{(2 \pi)^3} .
\end{equation}

Since the mass of the nucleus is -- especially in our case -- much larger 
than the energies of the two emitted leptons we can use the following 
simplification: The nucleus takes nearly no energy but balances all momenta.
The recoil energy of the daughter nucleus of mass $m_{\rm daught}$ is bound within the following limits:
\be \label{eq_rec_energy}
  0 \leq \erec = \frac{ (\vec p + \vec p_\nu)^2}{2 m_{\rm daught}} \leq \erec _{\rm , max} 
               = \frac{p^2_{\rm max}}{2 m_{\rm daught}} = \frac{E^2_0 + 2E_0 m}{2 m_{\rm daught}}.
\ee
Here we denote without indices the quantities of the electron and with index $\nu$ the ones of the neutrinos. Since we will reserve the notation $E$ for the 
kinetic energy of the \belec\ we denote its total energy with $E_{\rm e}$.
$E_0$ is the maximum kinetic energy the electron can obtain. 

Therefore we need to count the state density of the electron and the neutrino only
\begin{eqnarray}
\label{eq_ndensityall}
  \rho(E_{\rm e}, E_\nu, d\Omega_e, d\Omega_\nu)    
          & =  & \frac{dn_{\rm e}}{dE_{\rm e} d\Omega_{\rm e}} \cdot  \frac{dn_{\nu}}{dE_\nu d\Omega_\nu} 
          = \frac{V^2 \cdot p_{\rm e} \cdot E_{\rm e} \cdot p_\nu \cdot E_\nu}{(2 \pi)^6}\\
          & = &\frac{V^2 \cdot \sqrt{ E^2_{\rm e} - m^2} \cdot E_{\rm e} \cdot \sqrt{E^2_\nu - m^2(\nu_{\rm e})} \cdot E_\nu}{(2 \pi)^6}. \nonumber
\end{eqnarray}

The transition matrix element $M$ can be divided into a leptonic part,
\mlep, and a hadronic one, \mhad . Usually the 
coupling is written separately and expressed in terms of 
Fermi's coupling constant \gfermi\ and the Cabibbo angle $\Theta_{\rm C}$:
\begin{equation}
\label{eq_matrixelement}
  M = \gfermi \cdot \cos{\Theta_{\rm C}} \cdot \mlep \cdot \mhad
\end{equation}
For an allowed or superallowed decay like that in tritium, 
the leptonic part \mtwolep\ essentially results in the 
probability of the two leptons to be found at the nucleus, which is $1/V$ 
for the neutrino and $1/V \cdot F(E,Z+1)$ for the electron, yielding
\begin{equation}
\label{eq_matrixelementlep}
  \mtwolep = \frac{1}{V^2} \cdot F(E,Z+1) .
\end{equation}
  
The Fermi function $F(E,Z+1)$ 
takes into account the final
electromagnetic interaction of the emitted \belec\ with the
daughter nucleus of nuclear charge $(Z+1)$. 
The Fermi function is approximately given by
\cite{holzschuh}
\begin{equation}
  F(E,Z+1) = \frac{2\pi \eta}{1 - exp( -2\pi \eta )}
\end{equation}
with the Sommerfeld parameter $\eta = \alpha (Z+1) / \beta$.

The spin structure and coupling to the nuclear spin, as well as its ($\beta,\nu$)
angular correlation, is usually contracted into the nuclear matrix element.

For an allowed or super-allowed transition the hadronic matrix element 
is independent of the kinetic energy of the electron. Generally
it can be divided into a vector current or 
Fermi part ($\Delta I_{\rm nucl} = 0$) 
and into an axial current or Gamov-Teller part 
($\Delta I_{\rm nucl} = \pm 1$).
In the former case, the spins of electron and neutrino couple to $S=0$, in the latter
case to $S = 1$. Summing over spin states and averaging over the ($\beta,\nu$) angular
correlation factor 
\begin{equation} \label{eq_angular_correlation}
  1 + a \cdot (\vec \beta \cdot \vec \beta_\nu)
\end{equation}
(with the electron velocity $\beta = v/c$ and the neutrino velocity $\beta_\nu = v_\nu /c$), 
the hadronic matrix element for tritium is \cite{robertson}
\begin{equation}
\label{eq_matrixelementhad_tritium}
  |M^2_{\rm nucl}({\rm tritium})| = 5.55 
\end{equation}

The phase space density (\ref{eq_ndensityall}) is distributed over a surface in the two-particle phase space which is defined by a $\delta$-function 
conserving the decay energy. With this prescription, we can integrate (\ref{eq_fermisgoldenrule}) over the continuum states and get the partial decay 
rate into a single channel; for instance, the ground state of the daughter system with probability $P_0$:
\bea \label{eq_total_rate}
  \Gamma_0 = P_0 & \cdot & \int_{E_{\rm tot}, E_{\rm tot, \nu}, \Omega, \Omega_\nu} 
 \frac{G_{\rm F}^2 \cdot \cos^2\Theta_{\rm C} \cdot \mtwohad \cdot}{(2\pi)^5} \cdot F(E ,Z+1)\\ \nonumber
    & \cdot & \sqrt{ E^2_{\rm e} - m^2} \cdot E_{\rm e} \cdot  \sqrt{E^2_\nu - m^2(\nu_{\rm e})} \cdot E_\nu \cdot 
      \left( 1 + a \cdot (\vec \beta \cdot \vec \beta_\nu) \right)\\ \nonumber
   & \cdot & \delta(Q + m - E_{\rm tot} - E_{\rm tot, \nu} - \erec )~
     dE_{\rm tot}~ dE_{\rm tot, \nu}~ d\Omega~ d\Omega_\nu 
\eea
A correct integration over the unobserved neutrino variables in (\ref{eq_total_rate}) has to respect the 
($\beta,\nu$) angular correlation factor (\ref{eq_angular_correlation}), which enters the recoil energy  (\ref{eq_rec_energy}). 
The variation of \erec\ near the endpoint is tiny \cite{weinheimer93}. 
Even for the most sensitive tritium \bdec\ experiment, the upcoming KATRIN experiment, the variation of \erec\ 
over the energy interval of investigation (the last 25~eV below the endpoint) can be neglected and replaced by a constant value 
of $\erec = 1.72$~eV, yielding a fixed endpoint 
\cite{masood07}
\be \ezero = Q - \erec \quad .
\ee
We can then integrate over $E_{\rm tot, \nu}$ simply by fixing it through the $\delta$-function to the missing energy 
 $E_{\rm tot, \nu} = (\ezero - E)$: the difference between endpoint energy \ezero\ and 
kinetic energy $E$ of the \belec . 
Further integration over the angles yields through (\ref{eq_angular_correlation}) an averaged nuclear matrix element, as mentioned above. 
Besides integrating over the ($\beta, \nu$)-continuum, we have to sum over all other final states. It is a double sum, one over the 3 neutrino mass 
eigenstates \mnui\ with probabilities $|U^2_{\rm ei}|$, the other over all of the electronic final states of the daughter system with probabilities 
$P_{\rm j}$ and excitation energies \vj . The latter are caused by the sudden change of the nuclear charge and the different nuclear charge
of the daughter atom/molecule. They give rise to shifted endpoint energies. Introducing the definition 
\be
  \varepsilon := (\ezero - E),
\ee
the total neutrino energy now amounts  to $E_{\rm tot, \nu} = \varepsilon - \vj$. 
Rather than in the total decay rate, we are interested in its energy spectrum $\gamma = d\Gamma / dE$, 
which we can read directly from (\ref{eq_total_rate}) without performing the second integration over the $\beta$ energy. Written in terms of 
$\varepsilon$ and summed up over the final states it reads
\bea  \label{eq_betaspec}
  \gamma  =  & & \frac{G_{\rm F}^2 \cdot \cos^2\Theta_{\rm C}}{2 \pi^3}
              \cdot \mtwohad \cdot F(E ,Z+1)\\ \nonumber
             & \cdot & (\ezero + m - \varepsilon) \cdot \sqrt{(\ezero + m -\varepsilon)^2 - m^2} \\ \nonumber
             & \cdot & \sum_{\rm i,j} |U^2_{\rm ei}| \cdot P_{\rm j} \cdot (\varepsilon-\vj) 
                                                       \cdot \sqrt{(\varepsilon - V_{\rm j})^2 - \mtwonui}
                     \cdot \Theta(\varepsilon - \vj - \mnui). 
\end{eqnarray}
We directly see the validity of the definition of the average electron neutrino mass squared \mtwonue\ by (\ref{eq_define_mnue}), 
if the different neutrino mass states cannot be resolved experimentally.
The $\Theta$-function confines the spectral components to the physical sector $\varepsilon - \vj - \mnui >0$.
This causes a technical difficulty in fitting mass values smaller than the sensitivity limit of the data, 
as statistical fluctuations of the measured spectrum might occur which can no longer be fitted within the allowed physical parameter space. 
Therefore, one has to define a reasonable mathematical continuation of the spectrum into the region which leads to $\chi^2$-parabolas 
around  $\mtwonui \approx 0$ (see e.g. \cite{weinheimer93}). 
But one may equally well use formulas describing a physical model with the signature of a spectrum stretching beyond \ezero\ like tachyonic neutrinos
\cite{ciborowski99} (with the caution, of course, that one should not to jump to spectacular conclusions  from significant fit values $\mtwonui<0$  
instead of carefully searching for systematic errors in the data). 

Furthermore, one may apply radiative corrections to the spectrum \cite{repco83,gardner04}. However, they are quite small and would influence the 
result on \mtwonue\ only by few a percent of its present systematic uncertainty. 
One may also raise the point of whether possible contributions from right-handed currents might lead to measurable spectral anomalies 
\cite{stephenson98, ignatiev06}. It has been checked that the present limits on the corresponding right-handed boson mass 
\cite{pdg08, severijns06} rule out a sizeable contribution within present experimental uncertainties. 
Even the forthcoming KATRIN experiment will hardly be sensitive to this problem \cite{bonn06}.

The \belec s are leaving the nucleus on a time scale much shorter than the typical Bohr velocities of the shell electrons of the mother isotope. Therefore, the excitation probabilies of electronic states -- and of vibrational-rotational excitations in the case of molecules --
can be calculated in the so-called sudden approximation from the overlap of
the primary electron wave function $\Psi_0$ with the wave
functions of the daughter  ion $\Psi_{\rm f,j}$
\begin{equation} \label{eq_sudden_approximation}
  \vj = |\left< \Psi_0| \Psi_{\rm f,j} \right>| ^2.
\end{equation}

We will calculate the first excited electronic states for the case of a decaying tritium atom.
The tritium (i.e., hydrogen) wave function for the electronic ground state is:
\begin{equation}
  \Psi_0 = \Psi_{100}^{Z=1}(r, \vartheta, \phi) = \frac{1}{\sqrt{\pi a^3_0}} \cdot {\rm e}^{-r/a_0}, 
\end{equation}
with Bohr`s radius $a_0 = 4\pi \varepsilon_0/me^2$.
The final daughter atom is a $^3$He$^+$ ion. Therefore its wave functions $\Psi_{\rm f,j} = \Psi_{nlm}^{Z=2}$ 
are hydrogen-like functions with nuclear charge $Z=2$. 
Due to the orthogonality of the spherical harmonics $Y_{\rm lm}(\vartheta,\phi)$ the overlap intergral (\ref{eq_sudden_approximation}) can only be
non-zero for excited final states  $\Psi_{\rm f,j} = \Psi_{n00}^{Z=2}$.
For $Z=2$ the first 3 hydrogen-like wave functions are:
\bea
 \Psi_{100}^{Z=2} &=&  \sqrt{\frac{8}{\pi a^3_0}} \cdot {\rm e}^{-2r/a_0}\\
 \Psi_{200}^{Z=2} &=& \frac{1}{\sqrt{\pi a^3_0}} \cdot ( 1 - r/a_0) \cdot {\rm e}^{-r/a_0}\\ 
 \Psi_{300}^{Z=2} &=& \sqrt{ \frac{8}{27\pi a^3_0} } \cdot \left( 1 - \frac{4r}{3a_0} + \frac{8r^2}{27a^2_0} \right) 
                    \cdot {\rm e}^{-2r/3a_0}.
\eea
We can compute the overlap integral (\ref{eq_sudden_approximation}) using the following relation:
\begin{equation}
  \int_0^\infty r^n \exp{(-r/\mu)} ~dr = n! ~\mu^{n+1}\\
\end{equation}
The transition probability $P_0$ to the  $^3$He$^+$ ground state ($n=1$) amounts to:
\bea
  P_0 & = & \left| \int \int \int \frac{1}{\sqrt{\pi a^3_0}} \cdot {\rm e}^{-r/a_0} \cdot \sqrt{\frac{8}{\pi a^3_0}} \cdot {\rm e}^{-2r/a_0} \cdot r^2 
    \underbrace{\sin{\theta} d\theta d\phi}_{= 4\pi} dr \right|^2\\ \nonumber
      & = & \left| \frac{8 \sqrt{2}}{a^3_0} \cdot \int {\rm e}^{-3r/a_0} r^2 dr \right|^2\\ \nonumber
      & = & \left| \frac{8 \sqrt{2}}{a^3_0} \cdot \frac{2 a^3_0}{27}\right|^2\\ \nonumber
      & = & \left| \frac{16 \sqrt{2} }{27} \right|^2 = \frac{512}{729} = 0.702
\eea
The transition probability $P_1$ to the first excited state of the $^3$He$^+$ ion ($n=2$) is:
\bea
  P_1 & = & \left| \int \int \int \frac{1}{\sqrt{\pi a^3_0}} \cdot {\rm e}^{-r/a_0} 
                   \cdot \frac{1}{\sqrt{\pi a^3_0}} \cdot ( 1 - r/a_0) \cdot {\rm e}^{-r/a_0} \cdot r^2 
    \underbrace{\sin{\theta} d\theta d\phi}_{= 4\pi} dr \right|^2\\ \nonumber
      & = & \left| \frac{4\pi}{\pi a^3_0} \cdot \int  ( 1 - r/a_0) \cdot {\rm e}^{-2r/a_0} \cdot r^2 dr \right|^2\\ \nonumber
       & = & \left| \frac{4}{a^3_0} \cdot \left( 2 \frac{a^3_0}{8} - \frac{3 a_0^3}{8} \right) \right|^2\\ \nonumber
      & = & \left| \frac{-1}{2} \right|^2 = 0.25
\eea
Therefore, the first two electronic final states $P_0 + P_1$ comprise already more than 95~\% of all final states.
In addition to the excited states of the $^3$He$^+$ ion, there are also continuum states, which are more difficult to compute.
 
The excitation energies of the excited electronic states of the $^3$He$^+$ ion are:
\be
  V_{\rm n-1} = E(\Psi_{n00}^{Z=2}) - E(\Psi_{100}^{Z=2}) =  \frac{m (\alpha Z)^2}{2} \left( 1 - \frac{1}{n^2} \right) = 
               \left( 1 - \frac{1}{n^2} \right) \cdot 54.4~\rm{eV}.  
\ee
Thus, the excitation energy to the first excited level is $V_2 = 40.8$~eV. 

\begin{figure}[b!]
\centerline{\includegraphics[width=0.7\textwidth]{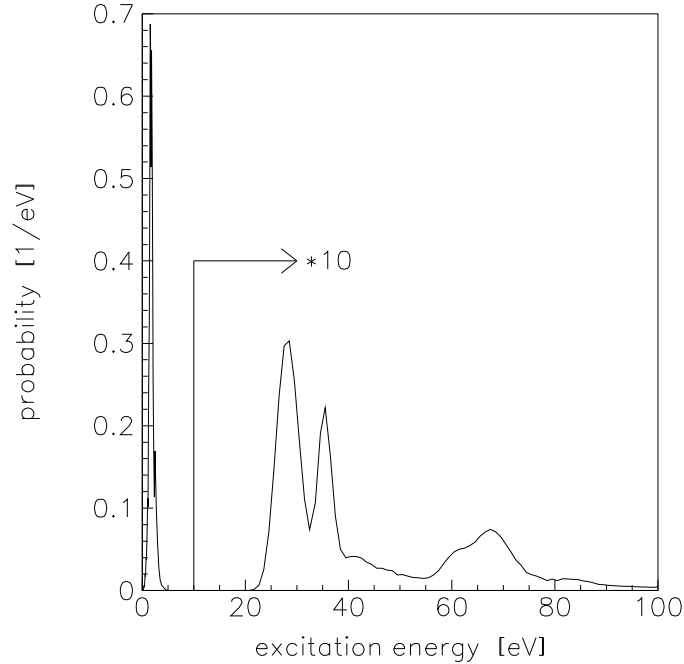}}    
\caption{Excitation spectrum of the daugther 
($^3$HeT)$^+$ in $\beta$ decay of molecular tritium \cite{saenz}.
} \label{fig_final_states}
\end{figure}

In practice, all tritium sources so far have been using molecular tritium sources, containing the molecule \ttwo .
The wave functions of the tritium molecule are much more complicated, since in addition to two identical electrons they comprise also the description of 
rotational and vibrational states, which will be excited during the \bdec\ as well.
Figure \ref{fig_final_states} shows a recent numerical calculation of the final states of the \ttwo\ molecule.
The transition to the electronic ground state of the $^3$HeT$^+$ daughter ion is not a single state, but broadened due to rotational-vibrational
excitation with a Gaussian standard deviation of $\sigma = 0.42$~eV. Secondly the first group of excitated states starts at around $V_{\rm j} = 25$~eV.
More recent calculations agree to these results \cite{doss08}.

\begin{figure}[t!]
\centerline{\includegraphics[width=0.9\textwidth]{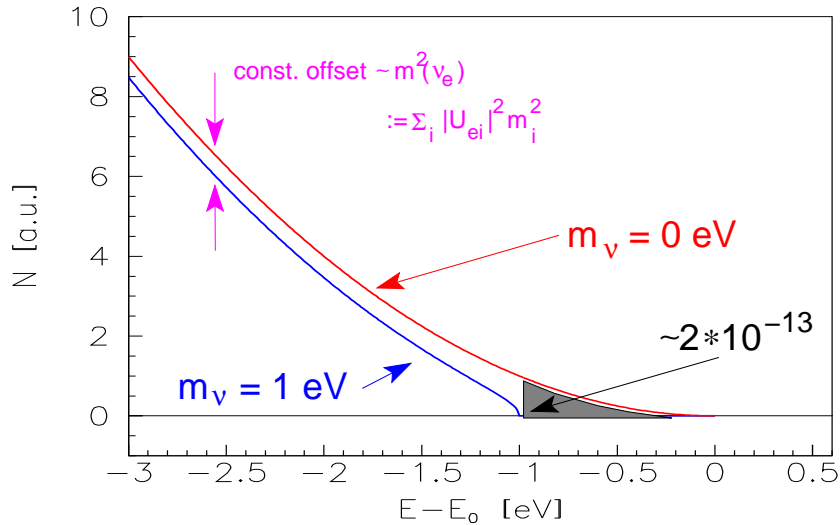}}
\caption{Expanded \bspec\ around its endpoint \ezero\
   for $\mnue = 0$ (red line) and for an arbitrarily chosen neutrino mass
  of 1~eV (blue line).
  In the case of tritium, the gray-shaded area corresponds to a fraction of $2 \cdot 10^{-13}$ of all
  tritium \bdec s.
\label{fig_beta_spec}}
\end{figure}

The neutrino mass influences the \bspec\ only at the upper end
below \ezero , where the neutrino is non-relativistic and can exhibit its massive character. 
The relative influence decreases in proportion to
$\mtwonue / \varepsilon^2$ (see figure \ref{fig_beta_spec}) leading far
below the endpoint to a small constant offset
proportional to $-\mtwonue$.

\begin{figure}[t!]
\includegraphics[width=0.33\textwidth]{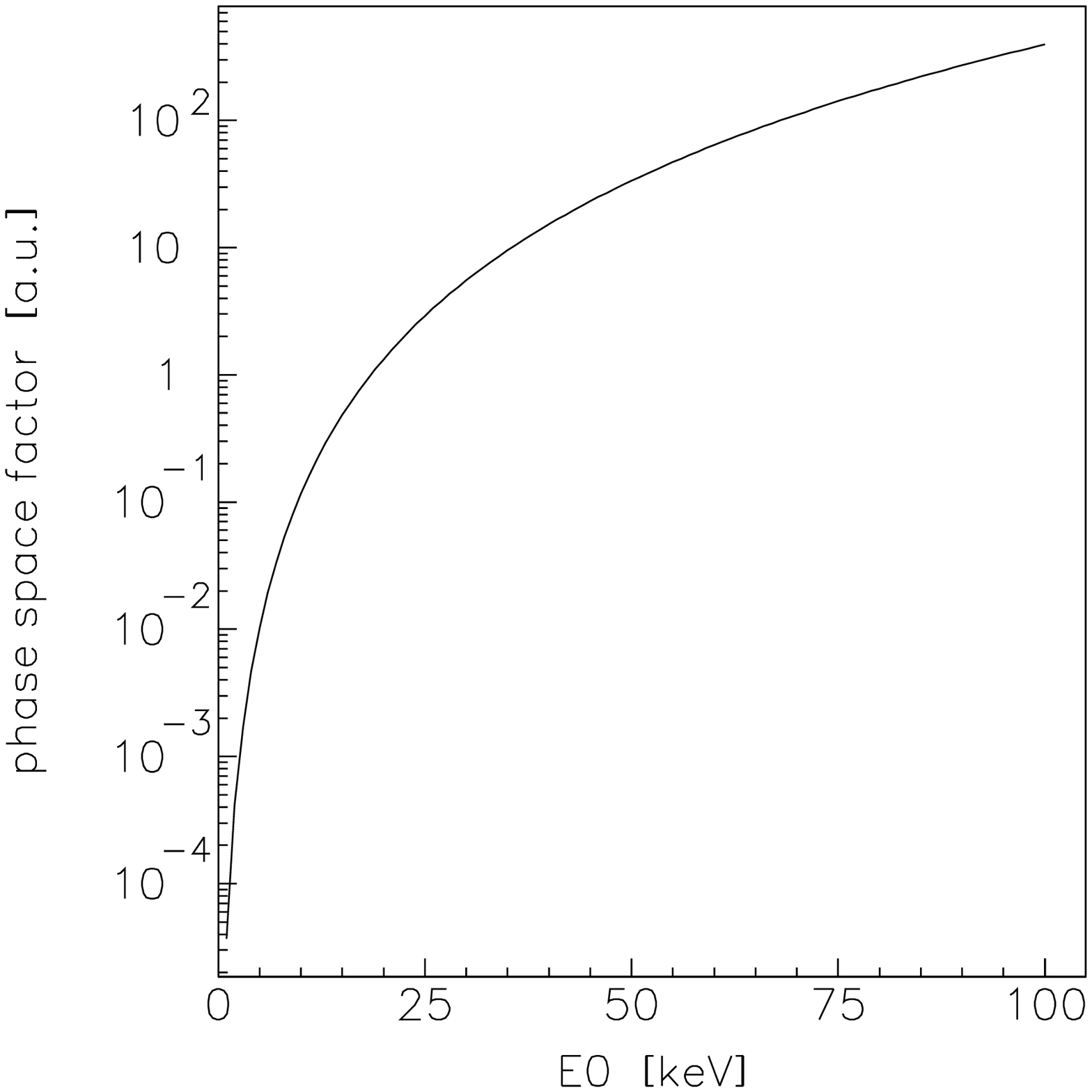}\includegraphics[width=0.33\textwidth]{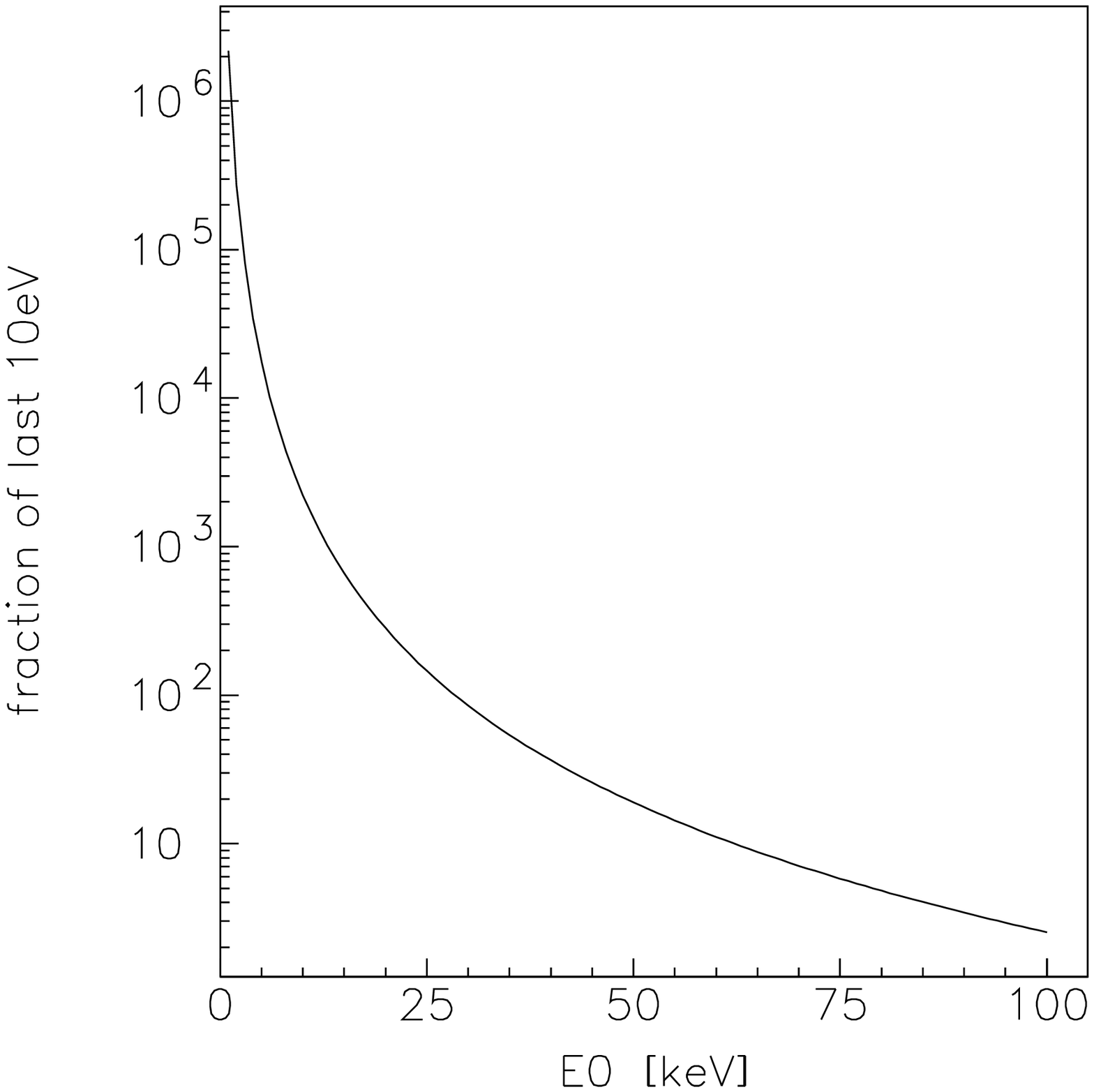}\includegraphics[width=0.33\textwidth]{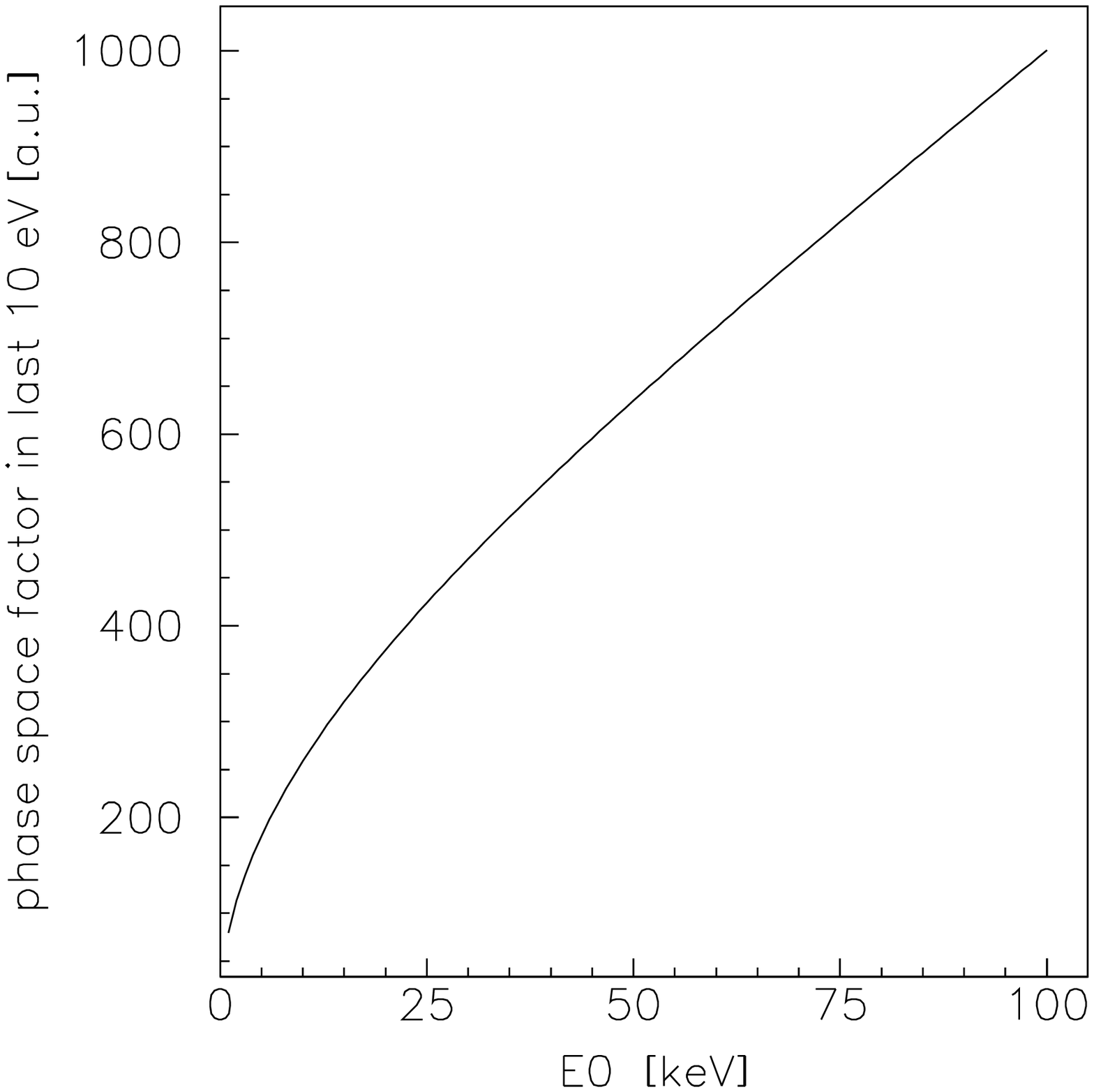}
\caption{Dependence on endpoint energy \ezero\ of total count rate (left), relative fraction in the last 10~eV below the endpoint (middle) and total count rate in the last 10~eV of a $\beta$ emitter (right).
  These numbers have been calculated for a super-allowed \bdec\ using (\ref{eq_betaspec}) for $\mnue = 0$ and neglecting possible final states as well as the Fermi function $F$.} 
\label{fig_qvalue_dependence}
\end{figure}

Figure \ref{fig_beta_spec} defines the requirements of a direct
neutrino mass experiment which investigates a \bspec : The task
is to resolve the tiny change of the spectral shape due to the
neutrino mass in the region just below the endpoint \ezero ,
where the count rate is going to vanish. Therefore, high energy
resolution is required combined with large source strength and
acceptance as well as low background rate.

Now we should firstly discuss, what is the best $\beta$ emitter for such a task. Figure \ref{fig_qvalue_dependence} shows the total count rate of
a super-allowed $\beta$ emitter as function of the endpoint energy. Of course, the total count rate rises strongly with \ezero , while the relative fraction 
in the last 10~eV below \ezero\ decreases. Interestingly, the total count rate in the last 10~eV below \ezero , which we can take as our energy region of interest
for determining the neutrino mass, is rather stable with regard to \ezero . From fig. \ref{fig_qvalue_dependence} one might argue, that the endpoint energy does not play a 
significant role in selecting the right $\beta$ isotope, but we have to consider the fact, that we need a certain energy resolution $\Delta E$ to determine the neutrino mass.
Experimentally it makes a huge difference, whether we have to achieve a certain $\Delta E$ at a low energy \ezero\ or at a higher one. Secondly, the \belec s of no interest
with regard to the neutrino mass could cause experimental problems ({\it e.g.} as background or pile-up) and again this arguement favors a low \ezero . 

Therefore, tritium is the standard isotope for this kind of study due to its low endpoint of 18.6~keV, its rather short half-life of 12.3~y,
its super-allowed shape of the \bspec ,
and its simple electronic structure. 
Tritium \bdec\ experiments using a tritium source and a separated 
electron spectro\-meter have been performed in search for the 
neutrino mass for more than 50~years. 

\rhenium\ is a second isotope suited to determine the neutrino mass. 
Due to the complicated electronic
structure of \rhenium\ and its primordial half life of $4.3 \cdot 10^{10}$~y the advantage of the 7 times lower
endpoint energy \ezero = 2.47~keV of \rhenium\ with respect to tritium can
only be exploited if the $\beta$ spectrometer measures  the entire
released energy, except that of the neutrino. This situation can
be realized by using a cryogenic bolometer as the $\beta$
spectrometer, which at the same time contains the $\beta$ emitter
\rhenium\ \cite{gatti_varenna} .

\section{Previous tritium neutrino mass experiments}

The majority of the published direct laboratory results on \mnue\
originate from
the investigation of tritium \bdec , while only two results from \rhenium\ have been
reported very recently (there are also results
from investigations of electron capture \cite{elec_capture} and bound state \bdec\
\cite{boundstate_bdec}, which are about 2 orders of magnitude less
stringent on the neutrino mass.).
In the long history of tritium $\beta$ decay experiments, about a dozen results
have been reported starting with the experiment
of Curran in the late forties yielding $\mtwonue < 1$~keV \cite{cur48}.

  In the beginning of the eighties a group from the Institute of Theoretical
  and Experimental Physics (ITEP) at Moscow \cite{itep} claimed the 
  discovery of a non-zero neutrino mass of around 30~\ev .
  The ITEP group used as $\beta$ source a
  thin film of tritiated valine combined with
  a new type of magnetic
  ``Tretyakov'' spectrometer.
  The first results testing the ITEP claim came from the experiments at the University of Z\"urich
   \cite{Zuericha} and  the Los Alamos National Laboratory (LANL)
  \cite{LANLa}. Both groups used similar Tretyakov-type
  spectrometers,  but more  advanced tritium sources with respect to the
  ITEP group.
  The Z\"urich group used a solid source
  of tritium implanted into carbon and later a self-assembling film
  of tritiated hydrocarbon chains.
  The LANL group developed a gaseous molecular tritium source avoiding
  solid state corrections.
  Both experiments disproved the ITEP result. The reason for the `mass signal' at
  ITEP was twofold: the energy loss correction was
  probably overestimated, and a  $^3$He--T mass difference
  measurement \cite{litmaa} confirming the endpoint energy of the
  ITEP result, turned out only later to be significantly wrong \cite{vanDyck,nagy06}.

\begin{figure}[tb]
\centerline{\includegraphics[width=0.75\textwidth]{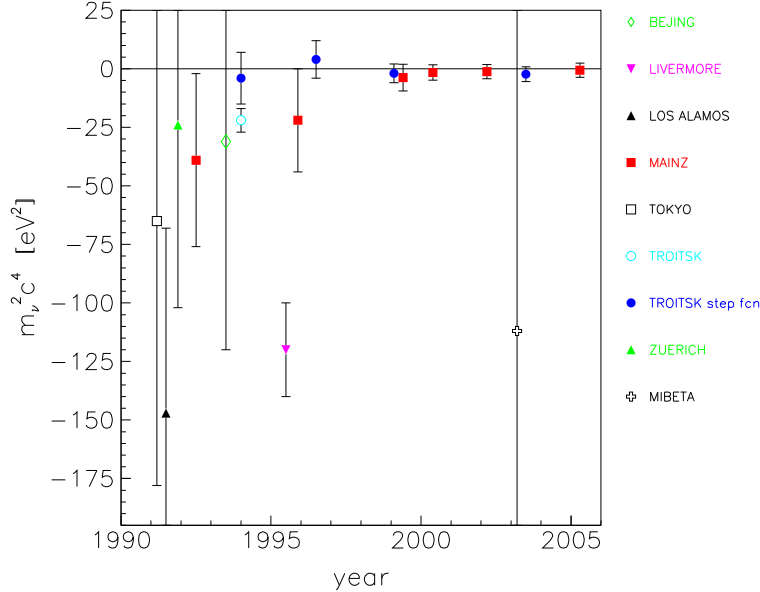}}
%\picplace{5cm}{2cm}
  \caption{Recent results of tritium \bdec\ experiments on the observable
  \mtwonue .The 
   experiments at Los Alamos, Z\"urich,
   Tokyo, Beijing and Livermore \protect\cite{LANLb,Zuerichb,Tokyo,Bejing,LLNL}
   used magnetic spectrometers, the tritium
   experiments at Mainz and Troitsk 
    \protect\cite{weinheimer99,kraus_05,belesev95,lobashev99} are using
   electrostatic spectrometers of the MAC-E-Filter type (see text).}
  \label{fig_tritiumexp}
\end{figure}

  Also in the nineties
  tritium \bdec\ experiments yielded controversially discussed results:
  Figure \ref{fig_tritiumexp} shows the final results of the
  experiments at LANL and Z\"urich together with the results from
  other more recent measurements with magnetic spectrometers
  at University of Tokyo, Lawrence Livermore National
  Laboratory and Beijing. The sensitivity
  on the neutrino mass have improved a lot but the values for the observable  \mtwonue\
  populated the unphysical negative \mtwonue\ region.
  In 1991 and 1994 two new
  experiments started data taking at Mainz and at Troitsk,  which used
  a new type of electrostatic spectrometer, so-called MAC-E-Filters,
  which were superior in energy resolution and luminosity with respect
  to the previous  magnetic spectrometers. However, even their early data were confirming the
  large negative \mtwonue\ values of the LANL and Livermore experiments
  when being analyzed over the last 500~eV of the \bspec\
  below the endpoint \ezero . But the large negative values of \mtwonue\ disappeared when analyzing only
  small intervals below the endpoint \ezero .
  This effect, which could only be investigated by the high luminosity MAC-E-Filters,
   pointed towards
  an underestimated or missing energy loss process, seemingly to be present
  in all experiments. The only common feature of the various experiment
  seemed to be the calculations of the electronic excitation energies
  and excitation probabilities of the daughter ions.
  Different theory groups checked these calculations in detail.
  The expansion was calculated to one order further and new interesting
  insight into this problem was obtained, but no significant changes
  were found \cite{saenz,doss08}.

  Then the Mainz group found the origin of the missing
  energy loss process for its experiment. The Mainz experiment
  used as tritium source a  film of molecular tritium
  quench-condensed onto aluminum or graphite substrates. Although the
  film was prepared as a homogenous thin film with flat surface, detailed studies
  showed \cite{fleischmann2} that the film undergoes a temperature-activated roughening transition
  into an inhomogeneous film by formation of microcrystals leading to
  unexpected large inelastic scattering probabilities.

  The Troitsk experiment on the other hand
  used a windowless gaseous molecular tritium source, similar to the LANL
  apparatus. Here, the influence of large-angle scattering of electrons
  magnetically trapped in the tritium source was not considered in the
  first analysis. After correcting for this effect
  the negative values for \mtwonue\ disappeared.

  The fact that more experimental results of the early nineties populate the region of negative \mtwonue\ values
  (see  fig. \ref{fig_tritiumexp})
  can be  understood by the following consideration \cite{robertson}:
  For $\varepsilon \gg \mnue $,  eq. (\ref{eq_betaspec}) can be expanded into
  \begin{equation}
    \frac{dN}{dE} \propto \varepsilon^2 - \mtwonue/2.
    \label{eq_betaspec_expanded}
  \end{equation}
  On the other hand the convolution of a \bspec\ (\ref{eq_betaspec}) with
  a Gaussian of width $\sigma$ leads to
  \begin{equation}
    \frac{dN}{dE} \propto \varepsilon^2 + \sigma^2.
    \label{eq_betaspec_sigma}
  \end{equation}
  Therefore, in the presence of
  a missed experimental broadening with Gaussian width $\sigma$ one expects a shift
  of the result on \mtwonue\ of
  \begin{equation}
    \Delta \mtwonue \approx - 2 \cdot \sigma^2,
    \label{eq_sigma_mtwonue}
  \end{equation}
  which gives rise to a negative value of \mtwonue\ \cite{robertson}.

\newpage
  \subsection{MAC-E-Filter} 

  The significant  improvement in the neutrino mass sensitivity by the Troitsk and the Mainz experiments
  are due to MAC-E-Filters (\underline{M}agnetic \underline{A}diabatic \underline{C}ollimation with an
  \underline{E}lectrostatic \underline{Filter}). This new type of spectrometer
  -- based on early work by Kruit \cite{kruit} -- was developed for the application to the tritium \bdec\
  at Mainz and Troitsk  independently \cite{pic92a,lob85}.
  The MAC-E-Filter combines high luminosity at low
  background and a high energy resolution, which are essential features
  to measure the
  neutrino mass from the endpoint region of a \bdec\ spectrum. 

  \begin{figure}[tb]
  \centerline{\includegraphics[angle=0,width=0.75\textwidth]{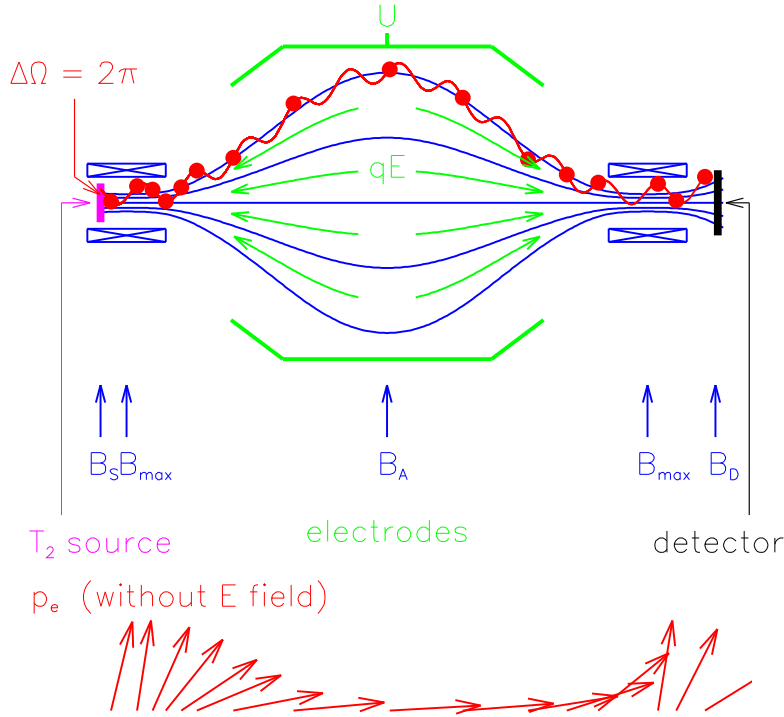}}
  \caption{Principle of the MAC-E-Filter. Top: experimental setup,
   bottom: momentum transformation due
   to adiabatic invariance of the orbital magnetic momentum $\mu$ in the
    inhomogeneous magnetic field.}
  \label{fig_mace}
  \end{figure}

  The main features of the MAC-E-Filter are illustrated in figure
  \ref{fig_mace}:
  two superconducting solenoids are producing a magnetic guiding field.
  The \belec s, starting from the tritium source
  in the left solenoid into the forward hemisphere, are
  guided magnetically on a cyclotron motion along the magnetic field lines
  into the spectrometer, thus resulting in an accepted solid
  angle of nearly $2 \pi$.
  On their way into the center of the spectrometer the magnetic
  field $B$ drops adiabatically by several orders of magnitude keeping
  the magnetic orbital moment $\mu$
  invariant  (equation given in non-relativistic approximation):
  \begin{equation}
    \mu = \frac{E_\perp}{B} = const.
  \end{equation}
  Therefore nearly all cyclotron energy $E_\perp$ is transformed into 
  longitudinal motion (see fig. \ref{fig_mace} bottom) 
  giving rise to a broad beam of electrons flying almost parallel to the
  magnetic field lines.

  This parallel beam of electrons is  energetically analyzed by
  applying an
  electrostatic barrier made up by a system of one or more cylindrical electrodes.
  The relative sharpness of this energy high-pass filter is only given by the ratio of the minimum
  magnetic field $B_{\rm min}$ reached at the electrostatic barrier in the so-called 
  analyzing plane 
  and the maximum magnetic field between
  \belec\ source and spectrometer $B_{\rm max}$:
  \begin{equation}
    \frac{\Delta E}{E} = \frac{B_{\rm min}}{B_{\rm max}}.
  \end{equation}

\begin{figure}[t!]
    \centerline{\includegraphics[angle=0,width=0.7\textwidth]{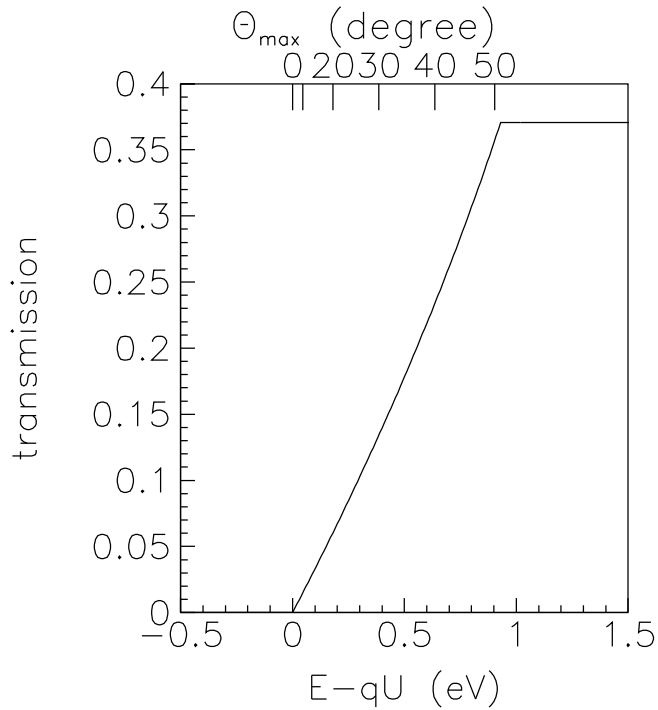}}
  \caption{Transmission function of the KATRIN experiment as function of the surplus energy $E-qU$. The KATRIN design settings \cite{katrin_tdr}
were used: $B_{\rm min} = 3 \cdot 10^{-4}$~T, $B_{\rm max} = 6$~T, $B_{\rm S} = 3.6$~T. The upper horizontal axis illustrates the dependence
of the maximum starting angle, which is transmitted at a given surplus energy. Clearly, electrons with 
larger starting angles reach the transmission condition later, since 
they still have a significant amount of cyclotron energy in the analysing plane at $B_{\rm min}$.}
  \label{fig_transmission_katrin}
\end{figure}

The exact shape of the transmission function can be calculated analytically. For an isotropically emitting source
of particles with charge  $q$ it reads: 
  \begin{equation}
     T(E,U) = \left\{ \begin{array}{ll} 
          0 & {\rm for~} E \leq qU\\
          1 - \sqrt{1 - \frac{E-qU}{E} \cdot \frac{B_{\rm S}}{B_{\rm min}}}
              & {\rm for~} qU < E < qU + \Delta E\\
           1 - \sqrt{1 - \frac{B_{\rm S}}{B_{\rm max}}}  & {\rm for~} E \geq qU + \Delta E 
     \end{array} \right.
  \end{equation}
We assume the electron source to be placed in a magnetic field $B_{\rm S}$ and that the retarding voltage of the spectrometer is $U$.
Fig. \ref{fig_transmission_katrin} shows the transmission functions for the settings of the KATRIN experiment (see section 4).

The \belec s are spiralling around the guiding magnetic field lines in zeroth approximation. Additionally, in non-homogeneous
electrical and magnetic fields they
feel a small drift $u$, which reads in first order \cite{pic92a} ($c=1$):
\begin{eqnarray}
  \vec u = \left( \frac{\vec E \times \vec B}{B^2} -\frac{(E_\perp + 2 E_{||})}{e\cdot B^3}
(\vec B \times \nabla_\perp \vec B) \right) 
\label{eq_drift}
\end{eqnarray}

The two recent tritium \bdec\ experiments at Mainz and at Troitsk 
use similar MAC-E-Filters with an energy resolution of 4.8~eV (3.5~eV) at Mainz (Troitsk).
The spectrometers differ slightly in size: 
the diameter and length of the Mainz (Troitsk)
spectrometer are 1~m (1.5~m) and 4~m (7~m). 
The major differences between the
two setups are the tritium sources:
Mainz uses as tritium source a thin film of molecular tritium quench-condensed 
on a cold graphite substrate, 
whereas Troitsk has chosen a windowless gaseous
molecular tritium source. After the upgrade of the Mainz experiment
in 1995-1997 both experiments ran with similar signal and similar
background rates.

\begin{figure}[t!]
    \centerline{\includegraphics[angle=0,width=\textwidth]{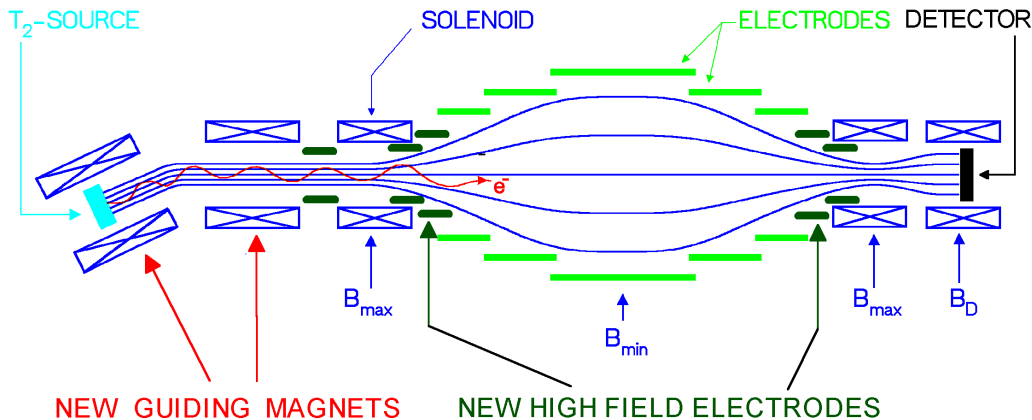}}
  \caption{The upgraded Mainz setup shown schematically. 
    The outer diameter amounts to 1~m, the distance
    from source to detector is 6~m.}
  \label{fig_mainz_newsetup}
\end{figure}

\subsection{The Mainz Neutrino Mass Experiment}

The Mainz setup was upgraded in 1995-1997 (see fig. \ref{fig_mainz_newsetup}), 
including the installation of 
a new tilted pair of superconducting
solenoids between  tritium source and spectrometer and the use of a 
new cryostat providing tritium film temperatures of below 2 K.
The first measure  eliminated source-correlated background and allowed 
the source strength to be increased significantly. The second measure
avoids the roughening transition of the homogeneously condensed tritium films with time
\cite{fleischmann2}, which previously gave rise to negative values of \mtwonue\ when the data
analysis used large intervals of the \bspec\ below the endpoint \ezero .
The upgrade was completed by the application
of HF pulses on one of the electrodes in between measurements every
20~s, and a full  automation of the apparatus and remote control. This former improvement
lowers and stabilizes the background, the latter one allows long--term measurements.

\begin{figure}[t!]
\centerline{\includegraphics[width=0.75\textwidth]{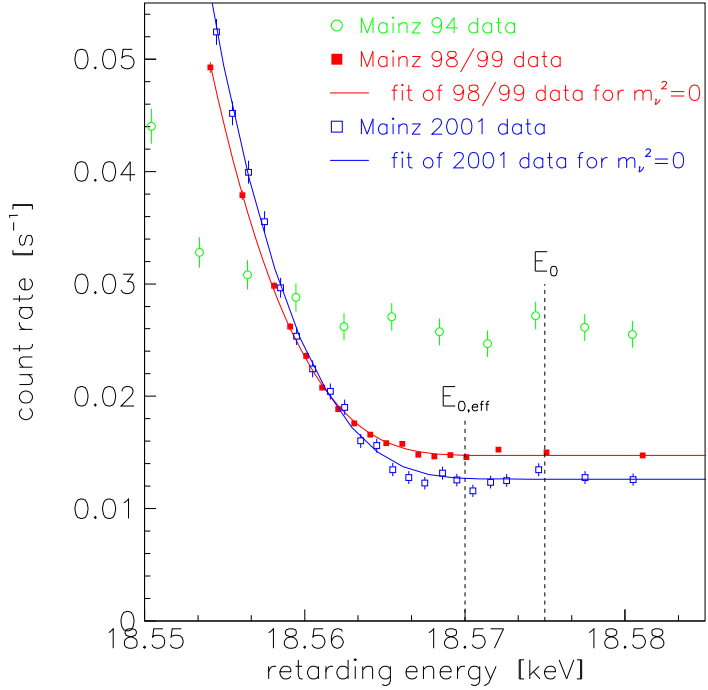}}
%\vspace*{-0.8cm}
\caption{Averaged count rate of the Mainz 1998/1999 data
(filled red squares) with fit for \mnue=0 (red line) 
and of the 2001 data (open blue squares) with fit for \mnue=0 (blue line) 
in comparison with previous Mainz data
from 1994 (open green circles)  
as a function of the retarding energy
 near the endpoint \ezero\  and effective endpoint $E_{0,{\rm eff}}$ (taking into account   
the width of the response function of the setup and 
the mean rotation-vibration excitation energy of the electronic 
ground state of the $\rm ^3HeT^+$ daughter molecule).}
\label{fig_mainz_data}
\end{figure}

Figure \ref{fig_mainz_data}
shows the endpoint region of the Mainz 1998, 1999 and 2001 data in comparison with the former
Mainz 1994 data. An improvement of the signal-to-background ratio
by a factor 10 by the upgrade of the Mainz experiment
as well as a significant enhancement of the
statistical quality of the data by longterm measurements are clearly visible.
The main systematic uncertainties of the Mainz experiment
are the inelastic scattering 
of \belec s within the tritium film,
the excitation of neighbor molecules due to sudden change of the nuclear charge
during \bdec , 
and the self-charging of the tritium film as a consequence of its radioactivity. 
As a result of detailed investigations in Mainz  
\cite{aseev,barth_erice97,bornschein03,kraus_05}
-- mostly by dedicated experiments -- 
the systematic corrections became
much better understood and their uncertainties were reduced
significantly.
The high-statistics Mainz data from 1998-2001 allowed the first
determination of the probability of the neighbor excitation
to occur in $(5.0 \pm 1.6 \pm 2.2)~\%$ of all \bdec s \cite{kraus_05} in good agreement
with the theoretical expectation \cite{kolos}.

The analysis of the last 70~eV below the endpoint of the 1998, 1999 and 2001 data,
resulted in \cite{kraus_05}
\begin{equation}
\mtwonue = (-0.6 \pm 2.2 \pm 2.1)~ \evtwo, 
\end{equation}
which corresponds to an upper limit  of
\begin{equation}
\mnue < 2.3~ \ev \quad {\rm (95~\%~C.L.)}
\end{equation}
This is the lowest model-independent upper limit of the neutrino mass obtained thus far.

\subsection{The Troitsk Neutrino Mass Experiment}

The windowless gaseous tritium source of the Troitsk experiment
\cite{lobashev99}  is essentially a tube of 5~cm diameter filled
with \ttwo\ resulting in a column density of $10^{17}$~molecules$\rm /cm^2$. 
The source is connected to the
ultrahigh vacuum of the spectrometer by a series a differential
pumping stations.

From their first measurement in 1994 onwards the Troitsk group has reported
the observation of a small, but significant
 anomaly in its experimental spectra starting a few eV below the
$\beta$ endpoint \ezero . This anomaly appears as a sharp step of
the count rate \cite{belesev95}. Because of the integrating property of the MAC-E-Filter, this step should correspond to a narrow line in the
primary spectrum with a relative intensity of about $10^{-10}$ of
the total decay rate. In 1998 the Troitsk group reported that the
position of this line oscillates with a frequency of 0.5 years
between 5~eV and 15~eV below \ezero\ \cite{lobashev99}. By 2000
the anomaly did no longer follow the 0.5 year periodicity, but
still existed in most data sets.
The reason for such an anomaly with these features is not clear.

In Mainz a similar behavior has been found only in one run taken under unfavorable conditions
\cite{kraus_05}. In dedicated measurements at Mainz, synchronously taken with the Troitsk experiment,
the anomaly was seen at Troitsk, but not at Mainz.
After some experimental inprovements the first two runs of 2001 at Troitsk
either gave no indication for
an  anomaly or only showed a small effect  with 2.5~mHz amplitude
 compared to the previous ones with amplitudes between 2.5~mHz
and 13~mHz. These findings as well as the Mainz data clearly support the assumption that the
Troitsk anomaly is due to an still unknown experimental artifact.

In presence of this problem, the Troitsk experiment is correcting for this 
anomaly by fitting an additional line to the \bspec\ run-by-run. 
Combining the 2001 results with the previous ones since 1994
gives \cite{lobashev03}
\begin{equation}
\mtwonue = (-2.3 \pm 2.5 \pm 2.0)~ \evtwo
\label{eq_troitsk_mtwo} 
\end{equation}
from which the Troitsk group deduces an upper limit of
\begin{equation}
\mnue < 2.05~ \ev \quad {\rm (95~\%~C.L.)}
\label{eq_troitsk_limit}
\end{equation}
The values of eq. (\ref{eq_troitsk_mtwo}) and (\ref{eq_troitsk_limit}) do 
not include the systematic uncertainty which is needed to be taken into
account when the timely-varying anomalous 
excess count rate at Troitsk is described run-by-run by an additional line.

\section{The Karlsruhe Tritium Neutrino Experiment KATRIN}

The previous Mainz and Troitsk experiment have reached their sensitivity limit on the neutrino mass with 2~\ev. Concerning a further and significant improvement on neutrino mass sensitivity,
the following lessons can be learned from these experiments:
\begin{itemize}
\item The MAC-E-Filter is a superior instrument to measure the endpoint region of the tritium \bspec\ with utmost sensitivity.
  Special care has to be taken of the background rate originating in the spectrometer.
\item The quench-condensed tritium source of the Mainz experiment is very well understood with small systematic uncertainties with regard to the Mainz sensitivity. All uncertainties
  could be improved except the self-charging of the tritium film, which causes an energy spread of 20~meV/monolayer of tritium.
  If this effect cannot be reduced or completely avoided, a significant further improvement is not possible using such a source.
\item The windowless gaseous tritium source at Troitsk which is based on the pioneer work at Los Alamos is complicated but served as a rather reliable source for a long time. 
  Special care has to be taken to stabilize the source to allow a long-term running and to avoid particle trapping. On the other hand such a windowless gaseous tritium
  source exhibits the smallest systematic uncertainties and would allow a significant improvement of the sensitivity on the neutrino mass. 
  To improve the luminosity and stability of such a windowless gaseous tritium source for a next generation of tritium \bdec\ experiment a large effort has to be taken. However,
  this seems to be achievable based on the strong expertise in tritium handling and purification, which is available in fusion technology. 
\end{itemize}

\begin{figure}[t]
\centerline{\includegraphics[angle=0,width=0.8\textwidth]{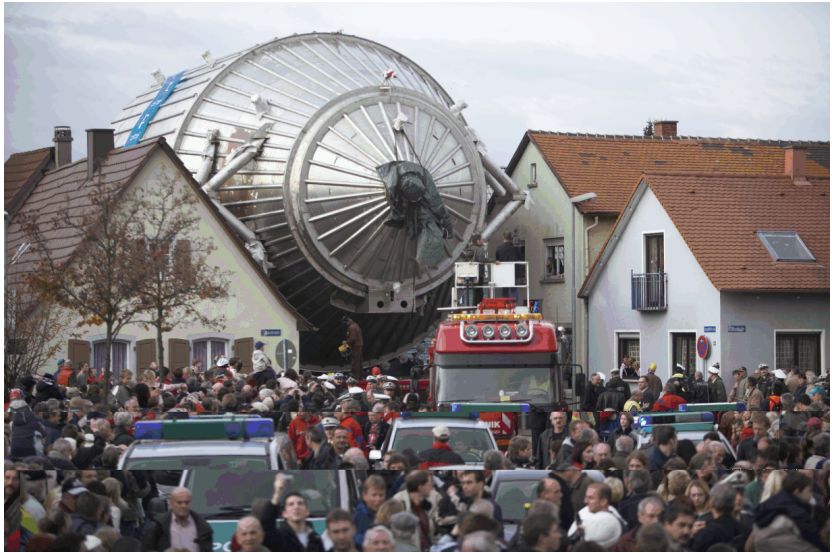}}
\caption{The KATRIN main spectrometer passes through the village of Leopoldshafen on its way from the river Rhine 
            to the Forschungszentrum Karlsruhe on November 25, 2006 (photo: FZ Karlsruhe).}
\label{fig_katrin_transport}
\end{figure}

Preliminary  ideas for next generation experiments on tritium \bdec\ in search for the absolute neutrino mass were presented  by the Troitsk 
\cite{lobashev_erice97} and Mainz \cite{barth_erice97} groups at a meeting in Erice in 1997. More details on the latter have been published by Bonn 
\etal\ \cite{bonn_macetof}. 
With the discovery of neutrino oscillations in 1998 \cite{superk_atmos98} the discussion gained momentum. 
Motivated by a long record in neutrino physics through the GALLEX and KARMEN experiments \cite{gallex,karmen}
and backed by the presence of a dedicated tritium laboratory on site, the Forschungszentrum Karlsruhe decided to get involved in the plans for a new neutrino mass experiment. 
It was named KATRIN standing for ``KArlsruhe TRItium Neutrino experiment''.

\begin{figure}[t]
\centerline{\includegraphics[angle=0,width=\textwidth]{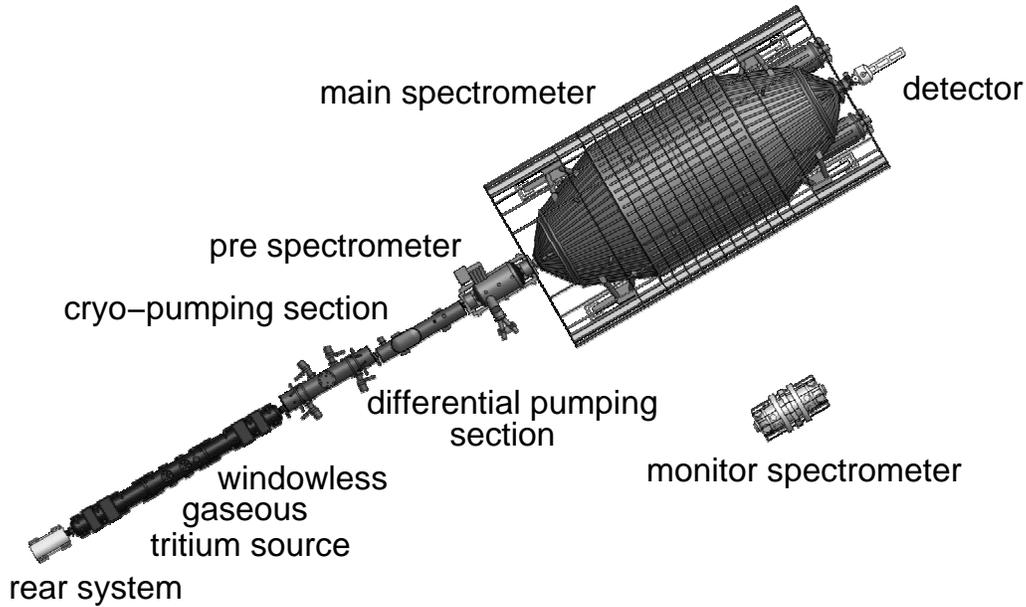}}
\caption{Schematic view of the 70~m long KATRIN experiment consisting of calibration and monitor rear system, 
windowless gaseous \ttwo -source, differential pumping and cryo-trapping section, small pre-spectrometer and large main spectrometer, 
segmented PIN-diode detector and separate monitor spectrometer.}\label{fig_katrin_setup}
\end{figure}

From a workshop in Bad Liebenzell in 2001  a letter of intent for the KATRIN experiment \cite{katrin_loi} emerged  
from close collaboration of group members from the earlier neutrino mass experiments at Los Alamos (now at University of Washington, Seattle and at
University of North Carolina, Chapel Hill), 
Mainz, and Troitsk with Forschungszentrum Karlsruhe. A design report \cite{katrin_tdr} was approved in 2004. 
Construction of the experiment is under way and expected to be completed  in 2011/12. 
The experiment aims for an improvement of the sensitivity limit by an order of magnitude down to to check the cosmologically relevant neutrino mass range 
and to distinguish degenerate neutrino mass scenarios from hierarchical ones. Furthermore,  Majorana neutrinos  
sufficiently massive to cause the neutrinoless double \bdec\ rate of $^{76}$Ge which part of the Heidelberg Moscow collaboration claims to have observed \cite{klapdor04}
would be observable in the KATRIN experiment in a model-independent way. 
The true challenge becomes clear by drawing attention to the experimental observable  
whose uncertainties have then to be lowered by two orders of magnitude.

Improving tritium $\beta$-spectroscopy by a factor of 100 evidently requires  brute force, based on proven experimental concepts. 
It was decided, therefore, to build a MAC-E-Filter with a diameter of 10 m, corresponding to a 100 times larger analyzing plane as compared 
to the pilot instruments at Mainz and Troitsk. 
Accordingly one gains a factor of 100 in quality factor which we may define as the product of accepted cross section of the source (``luminosity'') times the
resolving power $E/\Delta E$ for the emitted $\beta$-particles. 
Figure \ref{fig_katrin_transport} shows the spectrometer tank of KATRIN on its way to Forschungszentrum Karlsruhe, 
Figure \ref{fig_katrin_setup} depicts a schematic plan  of the whole 70 m long setup. 
Meanwhile the spectrometer has been set up and has reached its designed outgassing rate in the range of $10^{-12}$~mbar~l/(s~cm$^2$).

A decay rate of the order of $10^{11}$~Bq is aimed for in a source with a diameter of 9~cm. For the reason given above the KATRIN collaboration
decided to build a windowless gaseous tritium source (WGTS) in spite of its extraordinary demands in terms of size and cryo-techniques, 
which would be  required to handle the flux of $10^{19}$ \ttwo -molecules/s safely. \ttwo\
is injected at the midpoint of a 10~m long source tube kept at a temperature of 27~K by a 2-phase liquid neon bath. 
The integral column density of the source of $5 \cdot 10^{17}$~molecules/cm$^2$ has to be stabilized within 0.1~\%. 
Owing to background considerations, the \ttwo -flux entering the spectrometer should not exceed $10^5$ \ttwo -molecules/s. 
This will be achieved by differential pumping sections (DPS), followed by cryo-pumping sections (CPS) which trap residual \ttwo\ on argon frost at 4~K \cite{trap}. 
Each system reduces the throughput by $10^7$, which has been demonstrated for the cryo-pumping section by a dedicated experiment at Forschungszentrum Karlsruhe. 
The \ttwo -gas collected by the DPS-pumps will be purified and recycled.

\begin{figure}[t]
\centerline{
\includegraphics[angle=0,width=0.3\textwidth]{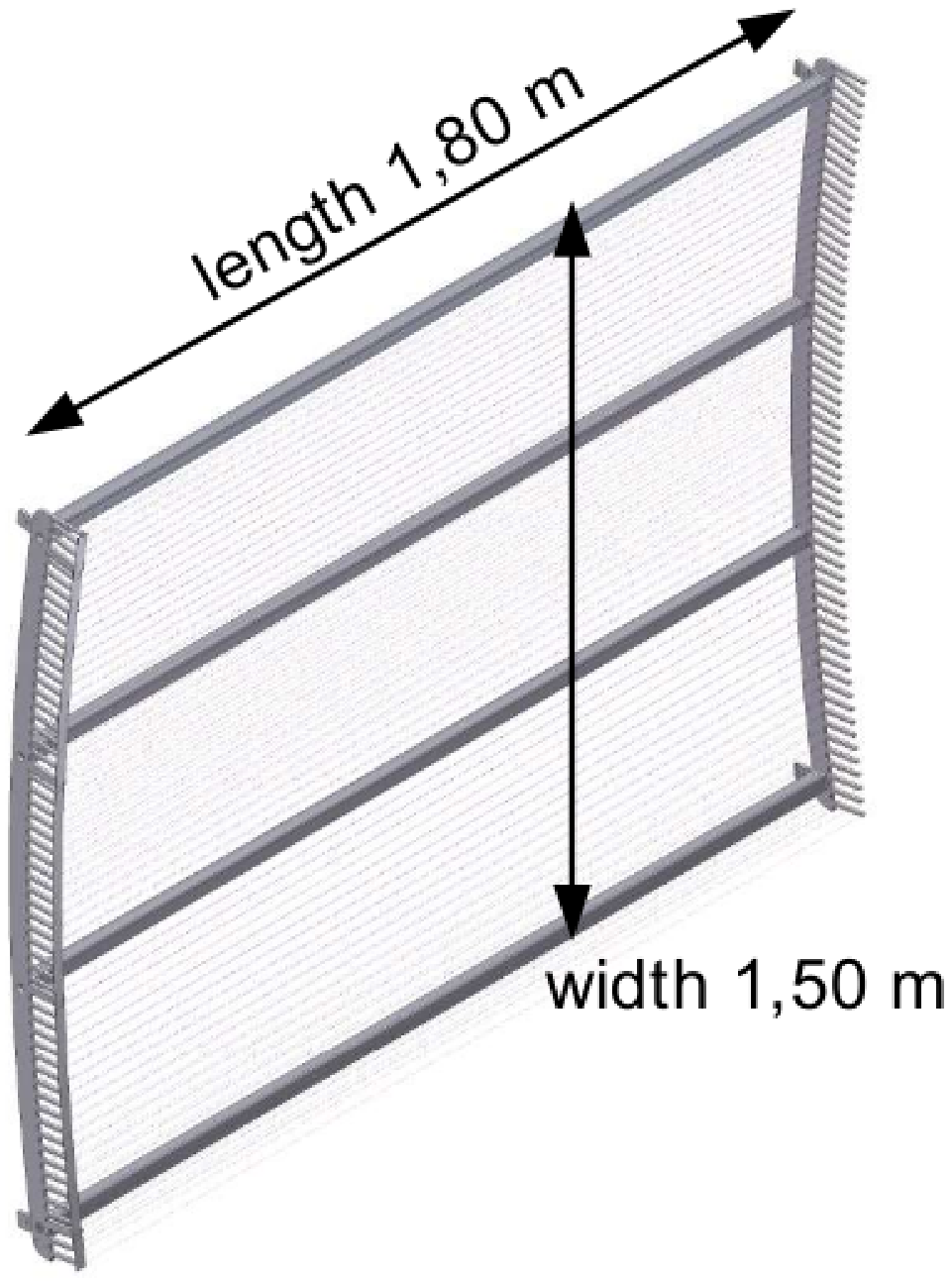} 
\hspace*{1cm}
\includegraphics[angle=0,width=0.6\textwidth]{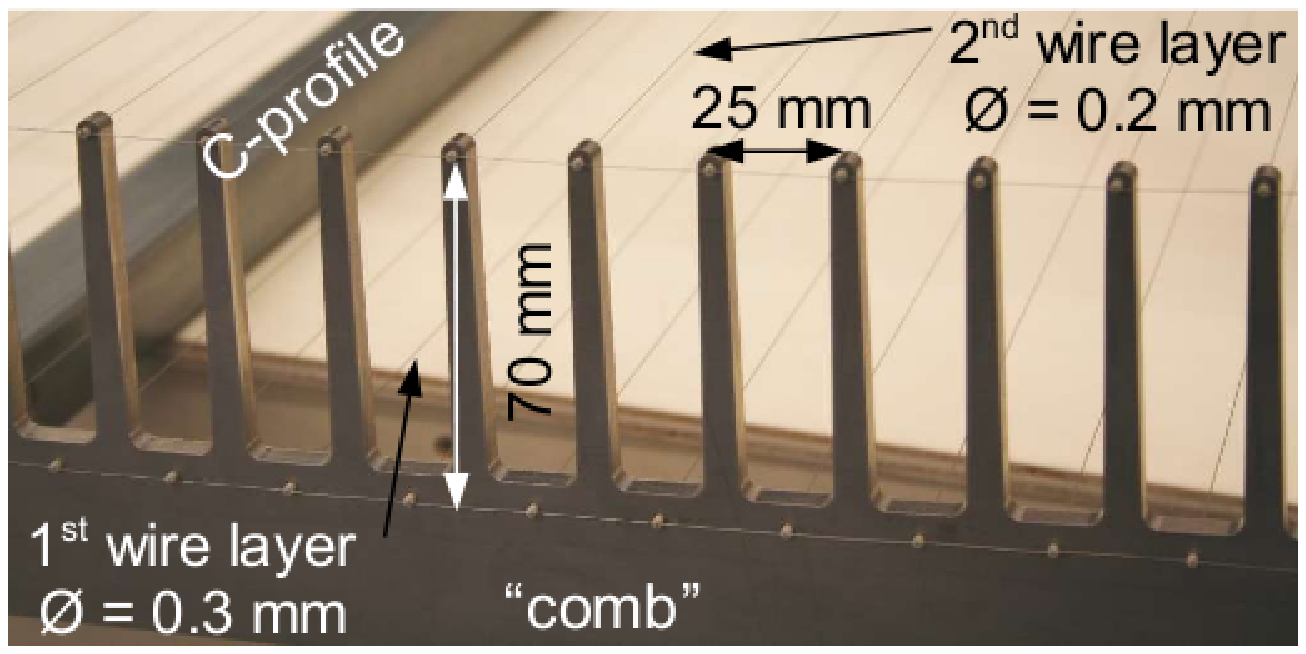}
}
\caption{Prototype of one of the 248 modules of the double-layer wire electrode system for the KATRIN main spectrometer. 
Wires with a diameter of 300~$\mu$m (200~$\mu$m) are used for the outer (inner) layer. The wires are mounted via precision ceramic holders 
onto a frame consisting of ``combs'' and C-profiles and keep their relative distance along their length within a few tenths of a mm. 
Materials are chosen to be non-magnetic and bakable at 350~$^\circ$C in order to reach the required low outgassing rate of $10^{-12}$~mbar~l/(s~cm$^2$) \cite{weinheimer06}.}
\label{fig_inner_electrode}
\end{figure}

A pre-spectrometer will transmit only the uppermost part of the \bspec\ into the main spectrometer in order to reduce the rate of background-producing 
ionization events therein. The entire pre- and main spectrometer vessels will each be put on their respective analyzing potentials, 
which are shifted within the vacuum tank by about -200~V, however, due to the installation of a background reducing inner screen grid system (fig. \ref{fig_inner_electrode}). 
A ratio of the maximum magnetic field in the pinch magnet over the minimum magnetic field in the central analyzing plane of the main spectrometer 
of 20000 provides an energy resolution of $\Delta E = 0.93$~eV near the tritium endpoint \ezero . 
The residual inhomogeneities of the electric retarding potential and the magnetic fields in the analyzing plane will be corrected by the 
spatial information from a 148 pixel PIN diode detector. 
Active and passive shields will minimize the background rate at the detector. 
Additional post-acceleration will reduce the background rate within the energy window of interest. 
Special care has to be taken to stabilize and to measure the retarding voltage. 
Therefore, the spectrometer of the former Mainz Neutrino Mass experiment will be operated at KATRIN as a high voltage monitor spectrometer 
which continuously measures the position of  the \kr -K32 conversion electron line at 17.8~keV, in parallel to the retarding energy of the main spectrometer. 
To that end its energy resolution has been refined to $\Delta E = 1$~eV.

The \belec s will be guided from the source through the spectrometer to the detector within a magnetic flux tube of 191~T~cm$^2$, 
which is provided by a series of superconducting solenoids. 
This tight transverse confinement by the Lorentz force applies also to the $10^{11}$ daughter ions per second, emerging from \bdec\ in the source tube, 
as well as to the $10^{12}$ electron-ion pairs per second produced therein by the \belec -flux through ionization of \ttwo\ molecules. 
The strong magnetic field of 3.5~T within the source is confining this plasma strictly in the transverse direction such that charged particles 
cannot diffuse to the conducting wall of the source tube for getting neutralized. 
The question how the plasma in the source becomes then neutralized or to which potential it might charge up eventually, 
has been raised and dealt with only recently \cite{nasti2005}.
The salient point is, however, that the longitudinal mobility is not influenced by the magnetic field. 
Hence the resulting high longitudinal conductance of the plasma will stabilize the potential along a magnetic field line 
to that value which this field line meets at the point where it crosses a rear wall. 
This provides a lever to control the plasma potential. Meanwhile the Troitsk group has performed a first experiment on the problem \cite{lobashev05}. 
They have mixed \kr\ into their gaseous \ttwo\ and searched for a broadening of the L$_{\rm III}$32-conversion line at 30.47~keV which might be due to an 
inhomogeneous source potential. Their data fit is compatible with a possible broadening of 0.2 eV, 
which would not affect their results but suggests further investigation at KATRIN. 
\begin{figure}[t!]
\centerline{\includegraphics[angle=0,width=0.5\textwidth]{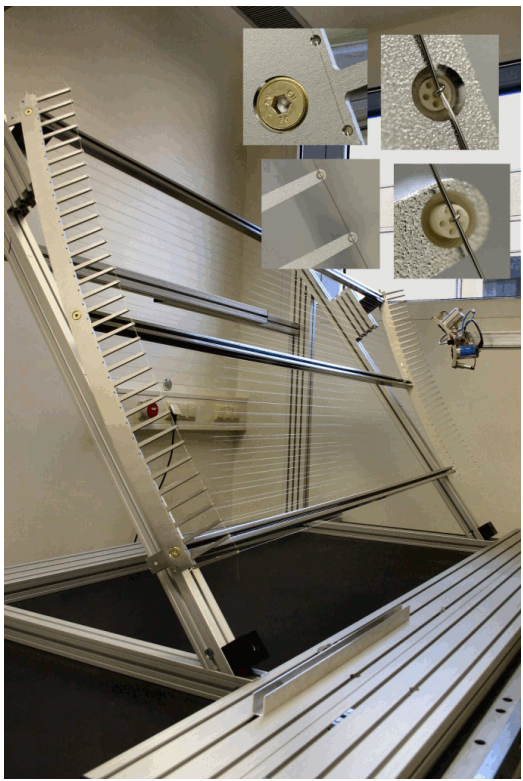}}
\caption{One of the final double-layer wire electrode modules on the 3-axis measurement table for quality assurance. 
The fixing of the wires inside the ceramics holders (see inserted smaller photos on the top right) with the connecting wires is checked with a high-resolution camera,
whereas wire position and wire tension are monitored by a specially developed 2-dim. laser sensor \cite{prall08}.}
\label{fig_inner_electrode_3dtable}
\end{figure}

\begin{figure}
\centerline{\includegraphics[angle=0,width=0.6\textwidth]{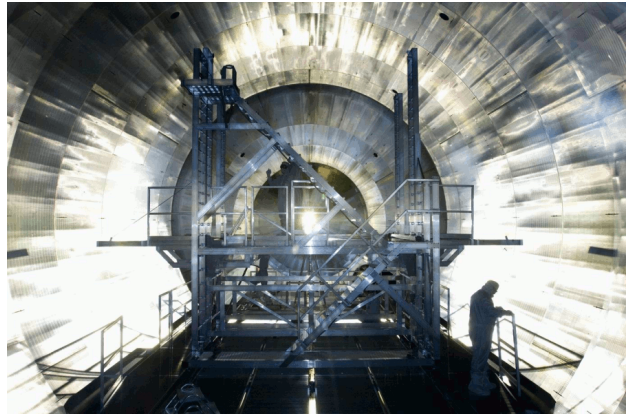}}
\caption{View into the KATRIN main spectrometer with scaffold installed. This scaffold was built by FZ Karlsruhe experts completely for the clean-room conditions of the 
 KATRIN main spectrometer (residual gas design pressure after out-baking: $10^{-11}$~mbar) \cite{luo07}. (photo: FZ Karlsruhe)} \label{fig_main_spec_scaffold}
\end{figure}

The sensitivity limit of KATRIN   has been simulated (see below) on the basis of a background rate of $10^{-2}$~cts/s, 
observed at Mainz and Troitsk under optimal conditions. Whether this small number can also be reached at the so much larger KATRIN-instrument 
-- or even be lowered -- has yet to be proven. On the one side, the large dimensions of the main spectrometer are helpful, 
as  they improve straight adiabatic motion due to reduced field gradients. 
On the other hand, the central flux tube faces a 100 times larger electrode surface at the analyzing potential from which secondary electrons might sneak in. 
Measurements at Mainz demonstrated that a large number of slow electrons at full potential emerge from the surface of the large central electrodes which are 
hit by cosmic muons and local radioactivity. 
But they are born outside the magnetic flux tube which crosses the detector; 
hence they are guided adiabatically past the detector. This decisive magnetic shielding effect was investigated at Mainz with an external $\gamma$-source, 
as well as by coincidence with traversing cosmic muons; 
a magnetic shielding factor of around $10^5$ was measured \cite{schall}. 
Furthermore similar checks at Troitsk pointed to 10 times better shielding \cite{lobashev_bg},
which probably results  from the better adiabaticity conditions of this larger instrument. 
In case  the axial symmetry of the electromagnetic field configuration is broken (e.g. by stray fields) 
the drift $u$ (\ref{eq_drift}), develops a radial component, which will be  all the faster the weaker the guiding field. 
This drift can transport slow electrons from the surface into the inner sensitive flux tube within which they are accelerated onto the detector. 
The effect is probably present at Mainz \cite{glueck}. 
After finishing tritium measurements in 2001, electrostatic  solutions were developed at Mainz, which strengthened shielding of surface electrons by an additional factor of $\approx 10$. 
This was achieved by covering the electrodes with negatively biased grids built from thin wires \cite{ostrick,flatt}.
Such grids are now under construction for the KATRIN-spectrometer (see fig. \ref{fig_inner_electrode},  \ref{fig_inner_electrode_3dtable}).
This measure (in addition to improved adiabaticity) will contribute decisively to keeping the background rate from this much larger instrument down to the design level of $10^{-2}$~cts/s
\cite{katrin_tdr}. The installation of wire electrode modules inside the KATRIN main spectrometer is a very challenging engineering task (see fig. \ref{fig_main_spec_scaffold}).

\begin{figure}[t!]
\centerline{\includegraphics[angle=0,width=0.7\textwidth]{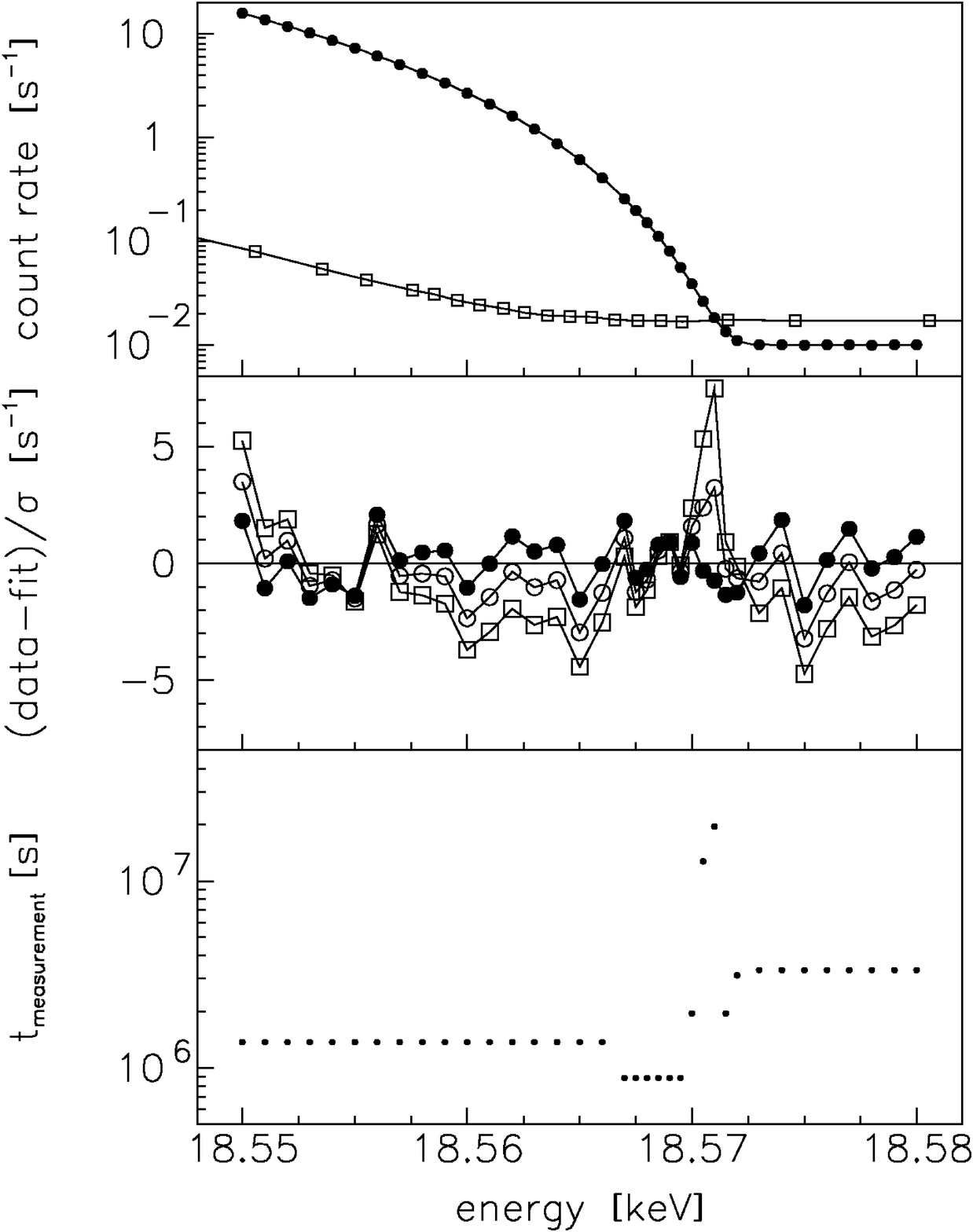}}
\caption{Top: Simulated $\beta$-spectra (assuming $\mnu = 0$ and $\ezero = 18.575$~keV) resulting from 3 years of KATRIN-running 
under KATRIN standard conditions (filled circles) and from phase 2 of the Mainz experiment for comparison (open squares). 
Middle: Difference of data and fit normalized to the statistical uncertainty for \mnue\ fixed in the fit to 0 \ev\ (filled circles), 0.35~\ev\ (open circles) and 0.5~\ev\  
(open squares). Bottom: Distribution of measuring points, optimized in position and measuring time.}\label{fig_katrin_mc}
\end{figure}

A simulated spectrum covering 3 years of data taking at KATRIN and is shown in fig. \ref{fig_katrin_mc}; a spectrum for the typcial measurement conditions at
Mainz is added for comparison. 
Due to the gain in the signal to background ratio, the region of optimal mass sensitivity around \ezero\ has moved much closer to the endpoint and one already  
notices at first glance  a marked mass effect for $\mnue = 0.5$~eV. 

One also notices that the typical third-power rise of the integral spectrum below \ezero\ is delayed. This is mainly due to rotational-vibrational excitations 
of the daughter molecule which centre at 1.72~eV and stretch up to more than 4~eV with a width of $\sigma_{\rm ro-vib} = 0.42$~eV (see fig. \ref{fig_final_states}).  
This width diminishes the mass sensitivity as compared to an atomic source with a sharp endpoint. 
At KATRIN this effect will be felt for the first time, but still amounts to only 5.5~\% sensitivity loss on \mtwonue, according to a simulation with 
standard KATRIN-parameters. 

\begin{figure}[t!]
\centerline{\includegraphics[angle=0,width=0.7\textwidth]{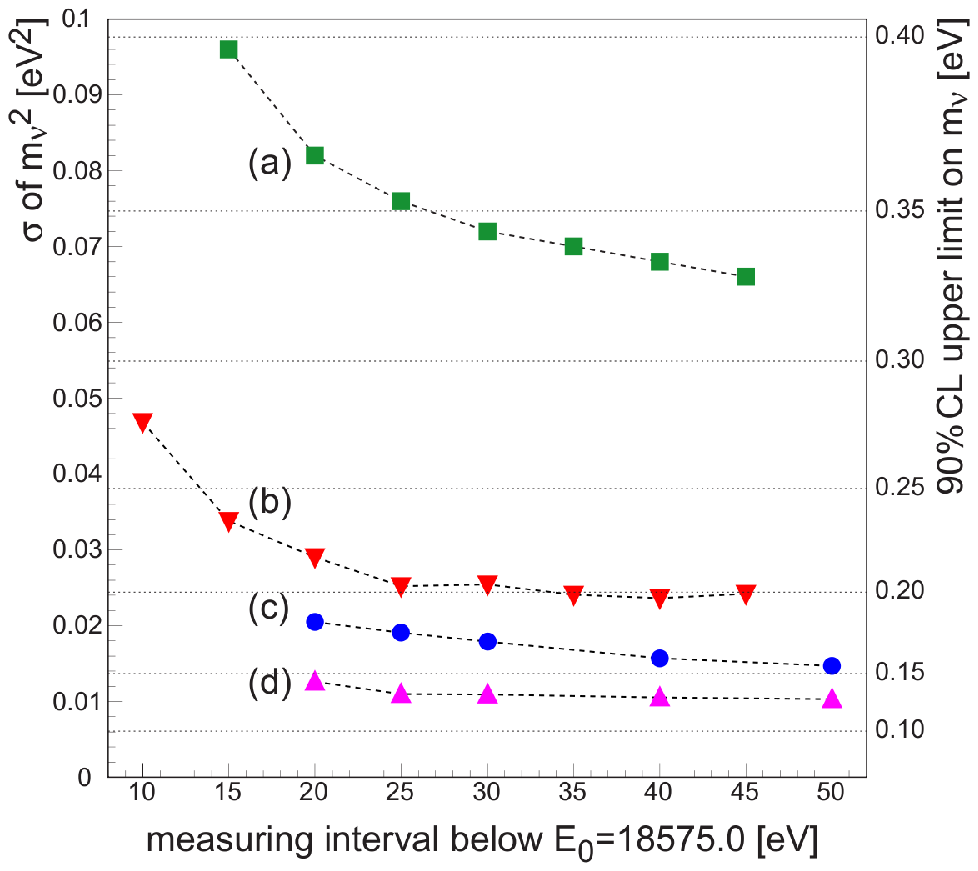}}
\caption{Simulations of statistical neutrino mass squared uncertainty expected at KATRIN after 3 years of running, 
calculated in dependence on the fit interval under following conditions. 
Spectrometer diameter = 7 m as originally proposed \cite{katrin_loi}: (a); final 10~m design \cite{katrin_tdr}: (b, c, d); background = $10^{-2}$~counts/s: 
(a, b, c); background = $10^{-3}$~counts/s: (d); equidistant measuring point distribution: (a, b); measuring point distribution optimized according to 
local mass sensitivity: (c, d) (reprinted from ref. \cite{katrin_tdr}).}\label{fig_katrin_stat_err}
\end{figure}
 
Fig. \ref{fig_katrin_stat_err} shows simulations of the statistical uncertainty of the observable 
and corresponding upper mass limits (without systematic uncertainties) which are expected from the 
KATRIN experiment after 3~years of data taking at background rates of $10^{-2}$~cts/s and $10^{-3}$~cts/s, respectively. 
They are plotted as a function of the width of the spectral interval, 
as measured with equidistant or optimized distribution of settings for analyzing potential as well as for measuring time. 
The dependence on the interval length is rather flat, 
in particular assuming a lower background. 
For the reference value  one expects to reach a total uncertainty $\Delta \mtwonue _{\rm stat}$ somewhat below $0.02~$~eV$^2$. 

Fortunately, the improved signal to noise ratio is very helpful with regard to the systematic uncertainties, 
as it allows to shorten the spectral interval under investigation below \ezero : Some of the systematic uncertaines decrease, others even vanish completely
as soon as the measurement interval drops below energy thresholds of inelastic processes, like the first electronic excitation of the ($^3$HeT)$^+$-ion at around 
25~eV (see fig. \ref{fig_final_states}) 
and the minimum energy loss of inelastic scattering on T$_2$-molecules of about 10~eV \cite{aseev}. 
From fig. \ref{fig_katrin_stat_err} it is clear that KATRIN aims at measuring intervals of about 25~eV below \ezero , 
for which the following systematic uncertainties and the corresponding counter-measures play a role:
\begin{itemize}
  \item Uncertainty of the energy-dependent cross section of inelastic scattering of \belec s on \ttwo\ in the windowless gaseous tritium source.\\
  Countermeasures: energy loss measurements with an electron gun as done in Troitsk \cite{aseev} analyzed by special deconvolution methods \cite{wolff08}.
  \item Fluctuations of the \ttwo\ column density in the windowless gaseous tritium source.\\
  Countermeasures: temperature and pressure control of the tritium source to the $10^{-3}$ level, 
   laser Raman spectroscopy to monitor the \ttwo\ concentration compared to HT, DT, H$_2$, D$_2$ and HD \cite{lewis08}.
  \item Spatial inhomogeneity of the transmission function by inhomogeneities of electric retarding potential and the magnetic field in the 
    analyzing plane of the main spectrometer.\\
    Countermeasures: spatially resolved measurements with an electron gun or, alternatively, with an \kr\ conversion electron source.
  \item Stability of retardation voltage \cite{kaspar04}.\\
   Countermeasures: a) measurement of HV with ppm-precision by a HV-divider \cite{thuemmler07} and a voltage standard; 
    b) applying the retarding voltage also to the monitor spectrometer, which continuously measures \kr\ conversion electron lines \cite{kaspar08,ostrick08,venos05,rasulbaev08}.
  \item Electric potential inhomogeneities in the WGTS due to plasma effects.\\
  Countermeasures: potential-defining plate at the rear exit of the WGTS; monitoring of 
   potential within WGTS possible by special runs with \kr /\ttwo -mixtures.
\end{itemize}

Each systematic uncertainty contributes to the uncertainty of  \mtwonue\ with less than $0.0075$~\evtwo, resulting in a total systematic uncertainty of 
$\Delta \mtwonue _{\rm sys} = 0.017$~\evtwo. 
The improvement on the observable \mtwonue\ will be two orders of magnitude compared to previous experiments at Mainz and Troitsk. 
The total uncertainty will allow a sensitivity on \mnue\ of 0.2~\ev\ to be reached. 
If no neutrino mass is observed, this sensitivity corresponds to an upper limit on \mnue\ of 0.2~\ev\ at 90~\%~C.L, or, otherwise, to evidence for
(discovery of)  a non-zero neutrino mass value at $\mnue = 0.3$~\ev\ (0.35~\ev) with $3\sigma$ ($5 \sigma$) significance. 
For more details we refer to the KATRIN Design Report \cite{katrin_tdr}. 

\section{Conclusion}

Among various ways to address the absolute neutrino mass scale
the investigation of the shape of \bdec\ spectra around the endpoint is the
only model-independent method. This direct method is complementary to the
search for the neutrinoless double \bdec\ and to the information from astrophysics
and cosmology.

The investigation of the endpoint spectrum of the tritium \bdec\ is still
the most sensitive direct method. The tritium \bdec\ experiments at Mainz and Troitsk have
been finished yielding upper limits of about 2~eV/c$^2$. The new KATRIN experiment is being set up
at the Forschungszentrum Karlsruhe by an international collaboration. To measure the tritium \bspec\ near the endpoint
with lowest systematic uncertainties and highest count rate, the KATRIN collaboration is setting up a) a windowless gaseous tritium
source with a factor 100 more count rate than previous experiments and b) a doublet of two spectrometers of MAC-E-Filter type, which is connected
to the windowless gaseous tritium source via a complex tritium elimination and electron transport chain. KATRIN's 
large main spectrometer has a 100 times larger cross section and a 5 times higher energy resolution compared to the previous tritium $\beta$ spectrometers. 
The background design value is based on active background reduction methods at the spectrometer (double layer screening wire electrode system) and at the 
electron detector (active and passive shielding, low activity materials). 
The systematic uncertainties of KATRIN will be well under control by many calibration and monitoring activities, as well as by virtue of the small energy
interval of interest below the endpoint reducing the influence of inelastic processes.
KATRIN will enhance
the model-independent sensitivity on the neutrino mass further by one order of magnitude down to 0.2~eV.

\section*{ACKNOWLEDGMENTS}
The author would like to thank all colleagues and friends from the KATRIN, Mainz and Troitsk collaborations for fruitful discussions. Among them he 
would like especially to name Ernst Otten (some 
material of this lecture originates
from the review article of him with the author \cite{otten08}), 
Jochen Bonn, Guido Drexlin and Ferenc Gl\"uck; as well as Kathrin Valerius for carefully reading and correcting these lecture notes. A very special thank
The work by the author for the KATRIN experiment is supported by 
the German Bundesministerium f\"ur Bildung und Forschung, the Deutsche Forschungsgemeinschaft and the University of M\"unster.

{}
\end{document}